\documentclass[fleqn,usenatbib]{mnras}

\usepackage{newtxtext,newtxmath}
\usepackage[T1]{fontenc}
\usepackage{ae,aecompl}

\usepackage[pdftex]{graphicx}	% Including figure files
\usepackage{amsmath}	% Advanced maths commands
\usepackage{amssymb}	% Extra maths symbols

\usepackage{epsfig}
\usepackage{graphicx}
\usepackage{natbib}
\usepackage{lscape}
\usepackage{threeparttable} 
\usepackage{xspace}
\usepackage{color}
\usepackage{epstopdf}
\usepackage{gensymb}
\usepackage{multirow}
\usepackage{booktabs}
\usepackage{times}
\usepackage[utf8x]{inputenc}
\usepackage{array}
\newcommand{\integral}{{\it INTEGRAL}}
\newcommand{\swift}{{\it Swift}}
\newcommand{\xrt}{{\it Swift}/XRT}
\newcommand{\bat}{{\it Swift}/BAT}

\newcommand{\xmm}{{\it XMM-Newton}}
\newcommand{\chandra}{{\it Chandra}}

\newcommand{\nustar}{{\it NuSTAR}}

\newcommand{\ergcm}{erg cm$^{-2}$ s$^{-1}$}

\newcommand{\ergs}{erg s$^{-1}$}

\newcommand\T{\rule{0pt}{2.6ex}}       % Top strut
\newcommand\B{\rule[-1.2ex]{0pt}{0pt}} % Bottom strut

\title[A comprehensive study of RX1804]{New insights on the puzzling LMXB 1RXS J180408.9-342058: the intermediate state, the clocked type-I X-ray bursts and much more}

\author[A. Marino et al.]{\thanks{E-mail: alessio.marino@unipa.it}
A. Marino$^{1,2,3}$, M. Del Santo $^{2}$, M. Cocchi$^4$, A. D'A\`i$^2$, A. Segreto$^2$, C. Ferrigno$^5$,
\newauthor
 T. Di Salvo$^1$, J. Malzac$^3$, R. Iaria$^1$, L. Burderi$^6$ \\ 
$^{1}$ Universit\'a degli Studi di Palermo, Dipartimento di Fisica e Chimica, via Archirafi 36,  I-90123 Palermo, Italy.\\
 $^2$INAF/IASF Palermo, via Ugo La Malfa 153, I-90146 - Palermo, Italy. \\
 $^3$IRAP, Universit\`e de Toulouse, CNRS, UPS, CNES, Toulouse, France.\\
 $^4$IAPS/INAF, via del Fosso del Cavaliere 100, 00133 Rome, Italy.\\
$^5$ISDC, Department of Astronomy, University of Geneva, Chemin d'Ecogia 16, CH-1290 Versoix, Switzerland.   \\
$^6$Universit\`a degli Studi di Cagliari, Dipartimento di Fisica, SP Monserrato-Sestu km 0.7, I-09042 Monserrato, Italy. \\
}

\date{Accepted XXX. Received YYY; in original form ZZZ}

\pubyear{2019}
\begin{document}
\label{firstpage}
\maketitle
\begin{abstract}
1RXS J180408.9--342058 is a low mass X-ray binary hosting a neutron star, which shows X-ray activity at very different mass-accretion regimes, from very faint to almost the Eddington luminosity. In this work, we present a comprehensive X-ray study of this source using data from the Neil Gehrels \swift\ Observatory,  \nustar\ and \integral/JEM-X. In order to follow the spectral evolution, we analysed the 2015 outburst using \swift\ data and three Nustar observations. Besides the canonical hard and soft spectral states, we identified the rarely observed intermediate state. This was witnessed by the appeareance of the accretion disk emission in the spectrum (at $kT_{\rm disk}$ $\sim$0.7 keV) and the simultaneous cooling of the hot corona. In addition, we also unveiled a hard tail above 30 keV in this state. In the hard state, a thermal Comptonization model with two seed photons populations ($kT_{\rm s,1}\sim 1.5$ keV and $kT_{\rm s,2}\sim 0.4$ keV, respectively) and a hot Comptonising plasma, represents the physically best motivated scenario to describe the data. We also estimated a reflection fraction below 20\% in all states, while no constraints on the inclination and only lower limits on the inner disk radius could be inferred. Finally, we studied a number of type-I X-ray bursts displayed from the source, one of them at the Eddington limit (observed with JEM-X). Their characteristics, combined with the clocked behaviour observed during the intermediate state, point out H/He composition for the accreted material, which makes unlikely the helium dwarf nature for the companion.
\end{abstract}

\begin{keywords}
accretion: accretion disks -- stars: neutron -- X-rays: binaries -- X-rays: bursts -- stars: individual (1RXS J180408.9-342058)
\end{keywords}

%%%%%%%%%%%%%%%%%%%%%%%%%%%%%%%%%%%%%%%%%%%%%%%%%%

%%%%%%%%%%%%%%%%% BODY OF PAPER %%%%%%%%%%%%%%%%%%

\section{Introduction}
Very Faint X-ray Transients (VFXTs) are a peculiar class of transients showing peak luminosities within 10$^{34}$ and a few 10$^{36}$ \ergs\  \citep[in the 2-10 keV;][]{Muno2005,Wijnands2006_GC}. A consistent number of this non-homogeneous class of transients have been classified as Low Mass X-ray Binaries (LMXBs) with neutron stars (NSs) \citep[e.g. ][]{Degenaar2009_vfxt,Degenaar2010,ArmasPadilla2013,VanDenEijnden2018}.   \\ These systems have been observed accreting at very low X-ray luminosities even over relatively long (months or years) time-scales \citep[e.g. ][]{DelSanto2007, IntZand2009, Degenaar2017_VFXT}. This behavior is associated with very low mass-accretion rate, although the physical origin is still an object of debate. Some of the proposed explanations involve: 1) a mass-accretion almost completely inhibited by a magnetic propeller effect \citep{Heinke2015,Degenaar2017_VFXT}; 2) the impossibility for a relatively big accretion disk to physically fit into the space between the two stars \citep[invoked for Ultra-Compact X-Ray Binaries, UCXBs,][]{King2006,Intzand2007,Hameury2016}; 3) low \citep{King2006} or high \citep[e.g. ][]{Muno2005} inclination angles; 4) the existence of a "period gap", as in Cataclismic Variables (CVs), where accretion via Roche lobes overflow is switched off but a low-level accretion via stellar wind is active \citep{Maccarone2013}. A subgroup of VFXTs is represented by the so-called "burst-only" sources. This class of NS-LMXBs were discovered from the long monitoring of the Galactic center with \emph{BeppoSAX} in the early 2000s \citep{Cocchi2001,Cornelisse2002}. Due to their low persistent luminosity, these sources were detected only during type-I X-ray burst events, i.e. thermonuclear explosions which occur on the NS surface due to the unstable ignition of the material accreted (see, for a recent review, \citealt{Galloway2017}). The type-I X-ray burst (hereafter burst) is a phenomenon of great astrophysical interest. In some cases a so-called Photospheric Radius Expansion (PRE) phase is observed when the Eddington luminosity is reached at the peak of the burst. The study of these bursts can be used as useful tool to infer the distance of NS-LMXBs \citep{Kuulkers2003} and, in some cases, also the mass and the radius of the NS \citep[see, for a review,][]{Ozel2016}. \\ Some of the "burst-only" systems, e.g. SAX J1806.5--2215, SAX J1753.5--2349, as well as 1RXS J180408.9--342058, have been also episodically displayed luminous outburst with peak luminosity above $10^{36}$ \ergs\ . These sources, observed at both "very faint" luminosity and standard luminosity regimes, are defined as "hybrid" VFXTs \citep[see, e.g.][]{Degenaar2009_vfxt,Delsanto2010,DelSanto2011_Atel}. \\
\subsection{1RXS J180408.9--342058}
1RXS J180408.9--342058 (hereafter RX1804) was discovered by \emph{ROSAT} in 1990 during the all-sky survey performed in the first half year of the mission \citep{Voges1999}. It was an unclassified X-ray source until 2012 April 16 when a burst was caught by JEM-X on board \integral\ and associated to the source \citep{Chenevez2012atel}. Thus, the source was classified as LMXB with NS showing a burst-only behaviour due to the very faint persistent luminosity (only an upper limit of 6 mCrab was inferred by JEM-X). Assuming that the burst reached the Eddington limit at the peak, \citep{Chenevez2012atel} estimated an upper limit to the distance of $\sim$5.8 kpc and to the persistent luminosity of roughly 10$^{35}$ \ergs\ . From the analysis of the same burst, with a different method for the peak flux estimation, \cite{Chelovekov2017} derived an upper limit for the distance to the source of 9.7$\pm$1.6 kpc. \\ 
On January 2015, the Burst Alert Telescope (BAT) on board the Neil Gehrels \swift\ Observatory (hereafter \swift) detected the source at $\sim$40 mCrab  \citep{Krimm2015}, revealing RX1804 to be a "hybrid" VFXT. Then,  a number of follow-up observations were triggered, i.e. with {\it MAXI/GSC} \citep{Negoro2015}, \xrt\ \citep{Krimm2015b} and \integral\  \citep{Boissay2015}, among the others. During the 4.5 months long outburst, the source exhibited both hard and soft states. \citet{Ludlam2016} performed a broad-band spectral analysis of the source between 0.45 and 50 keV using \emph{NuSTAR} and \xmm\ observations taken during the hard state. These authors modeled the reflection spectrum with \textsc{relxill} \citep{Garcia2014}, which allowed them to estimate an inclination $i\sim$\,$18{^\circ}-29{^\circ}$ and a strict upper limit on the inner disc radius ($\sim$\,22 km). \citet{Parikh2017_vhard} reported on unusual very hard spectra observed in three LMXBs with NS, including RX1804. In a number of XRT spectra (0.5-10 keV) they measured a very low value of the power-law photon index ($\Gamma \sim$\,1), significantly lower than the typical spectral indices observed in these sources in the hard state ($\sim$1.5-2).\\
The soft state was studied by \citet{Degenaar2016} using {\it NuSTAR}, \swift\ and \chandra\ observations obtained around the X-ray peak of the outburst. From the study of the reflection component, the authors reported on a broad iron line, signature of a disk extending close to the NS. Using the reflection model \textsc{reflionx}, an inner disk radius lower than 17 km and an inclination of the system around 30$^\circ$ were inferred. \citet{Baglio2016} performed a detailed near-infrared/optical/UV study of the source, with the aim of classifying the companion star. The lack of H and He I emission lines was considered by these authors as an evidence for an out-of-the-main-sequence secondary star. However, the marginal detection of the He II line would indicate the presence of He in the disk, possibly related to a helium white dwarf nature of the companion. Under this hypothesis, the authors classified the system as ultra-compact X-ray binary (UCXB) with an orbital period of about 40 min.  \\
In this paper we present a study of the temporal and spectral evolution of RX1804 during the bright 2015 outburst. Taking advantage of the almost continuous monitoring performed by \swift, we were able to individuate the two main hard and soft states and, for the first time for this source, the intermediate state, rarely observed in NS-LMXBs. In order to study the reflection component, we combined the three \nustar\ data archival observations with XRT and BAT. Two of them have already been reported in previous works (the hard state by \cite{Ludlam2016} and the soft state by \cite{Degenaar2016}). The \nustar\ observation performed during the intermediate state is reported in this paper for the first time. Furthermore, we report on results from the unpublished type-I X-ray burst observed by \integral\ in 2012 and the bursts detected during the Nustar intermediate state. 

\section{Observations and Data Reduction}
In the present work, we used data from several X-ray telescopes, such as \xrt\ , \bat\ and \nustar\ to study the evolution of the 2015 outburst and \integral/JEM-X, \nustar\ and \xrt\ for the type-I X-ray bursts study. 

\subsection{\swift}
The 2015 outburst was monitored by XRT, between MJD 57059 (February 2nd) and MJD 57166 (May 24th), with 24 pointings for a total exposure time of $\approx$31 ks. All these observations, with obsID: 00324360, 006300470 and 0008145, were performed in Window Timing (WT) mode. \\
XRT data were first reprocessed with the task \textsc{xrtpipeline}, included in the software package \textsc{HEASOFT} (v. 6.25). The source extraction procedure from the output cleaned event files was performed with \textsc{ds9}. A circular region of radius 20 pixels centered on the coordinates of the source was used, unless the count-rate was above 100 cts/s, for which we adopted an annulus region. In our observations we found 4 pointings, i.e. obsID 0032436029-32, affected by pile-up, for which we used an annulus region with an inner radius of 6 pixels. \\
Spectra and light curves of the source were extracted running the task \textsc{xrtproducts} using the source regions built in the previous step. The light curve of each observation was checked in order to individuate any flare and/or burst. On the 25 analyzed observations, we found and time-filtered 5 bursts. Each spectrum was rebinned with \textsc{grppha} in order to have 100 counts per bin, which allows the use of the $\chi^2$ statistics. \\  The BAT survey data, from MJD 57060 to MJD 57165, retrieved from the \textsc{HEASARC} public archive, were processed using \textsc{BAT-IMAGER} software \citep{Segreto2010}. This code, dedicated to the processing of coded mask instrument data, computes all-sky maps and, for each detected source, produces light curves and spectra. We extracted spectra with logarithmic binning and we used the official BAT spectral redistribution matrix.
\subsection{NuSTAR}
\label{ss:nustar}
\nustar\ observed RX1804 three times in 2015: on March 5 for $\sim$34 ks exposure time, on April 1 for $\sim$23 ks, and on April 14 for $\sim$20 ks (see Table \ref{tab:nustar}). The analysis of the second observation has never been reported before.\\
Data were reduced using the standard \textsc{Nustardas} task, incorporated in \textsc{Heasoft} (v. 6.25). We extracted high scientific products (light curves and spectra) using a circular area of 110" radius centered at R.A.\,=\,18:04:08.306 and Dec.\,=\,-34:20:50.63 as source region. In order to take into account any background non-uniformity on the detector, we extracted the background spectra using four circles of $\sim$50" radii in different positions with negligible contamination from the source. We then employed \textsc{Nuproducts} to build spectra and light curves. Five and ten type-I X-ray bursts were observed in the light curve of the source in the first and second observation, respectively. \\
We time filtered the spectra by removing the time intervals around these bursts; depending on the burst intensity and profile this resulted in a reduced exposure of 36.6 ks and 20.6 ks respectively. \\ 
We used data from both the two hard X-ray imaging telescopes on board \nustar, i.e. the focal plane mirror (FPM) A and B, and grouped the spectra by means of \textsc{Grppha} with a minimum of 150 counts per bin. This grouping was chosen taking into account the high number of counts from the source and the relatively poor \nustar\ spectral resolution, in order to avoid the oversampling of the energy response, as also done by other authors \citep[see, e.g.][]{Lanzuisi2016,Basak2017}. Furthermore, we did not sum the two spectra but rather fitted them together by leaving a floating cross-normalization constant as suggested by the \nustar\ team  for bright sources\footnote{On the FAQ page, issue 19: \url{https://heasarc.gsfc.nasa.gov/docs/nustar/nustar_faq.html}}. \\

\begin{table}
    \centering
    \begin{tabular}{l l l l l}
         \hline
         \hline
         ObsID & \multicolumn{2}{c}{Start Time} & Exposure & Ref.  \\
         & (UTC) & (MJD) & ks & \\
         \hline
         80001040002 & 2015-03-05 & 57086.4 & 37.3 & L16\\
         80001040004 & 2015-04-01 & 57113.7 & 23.4 & This work \\
         90101003002 & 2015-04-15 & 57126.5 & 20.2 & D16\\
         \hline
         \hline
    \end{tabular}
    \caption{List of the \nustar\ observations of RX1804 used in this work. L16 = \citet{Ludlam2016}, D16 = \citet{Degenaar2016}.}
    \label{tab:nustar}
\end{table}

\subsection{\integral\ }
\integral\ observed the 2012 type-I X-ray burst during the Science Window 116100100010, MJD 56033.350506, with exposure time of $\sim$1800 s. From this science window we extracted 2 s light curves of the burst in the 3-20 keV energy range using data from both the JEM-X monitors, i.e. JEM-X1 and JEM-X2, with the \integral\ data analysis software, \textsc{OSA11}.

\begin{figure*}
\centering
\includegraphics[height=21. cm, width=13 cm]{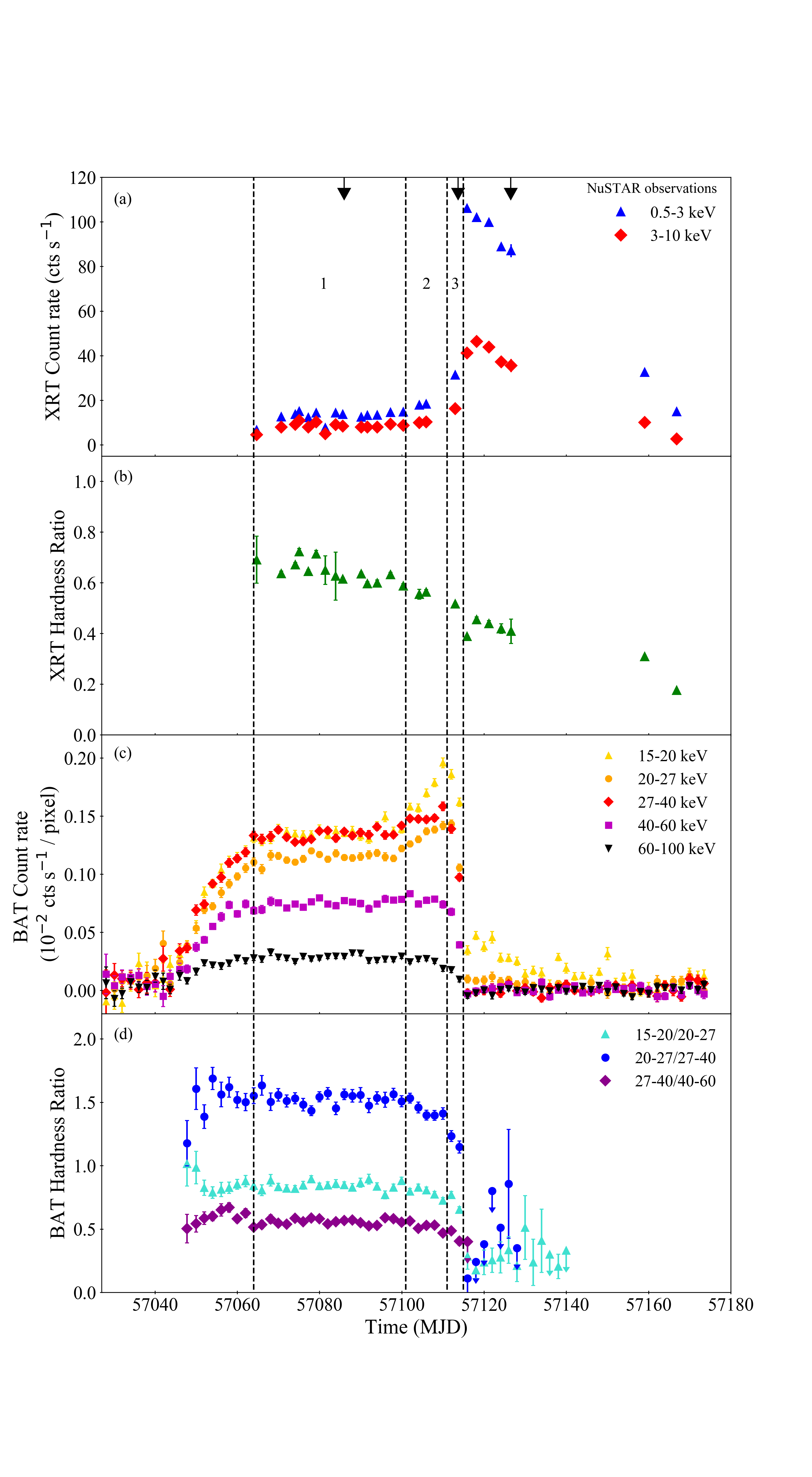}
\caption{ XRT and BAT light curves (subpanels (a) and (c), respectively) and hardness ratios (subpanels (b) and (d), respectively) of RX1804 during the 2015 outburst. The time intervals chosen for the first three BAT spectra used in this work are identified by dashed lines and labeled with numbers from 1 to 3. Onto the XRT light curve, the times of the three \nustar\ observations are flagged by arrows. In the BAT HR (d), points with large error bars are not shown for clarity.}
\label{fig:swift}
\end{figure*}

\section{The 2015 outburst}
\subsection{Data Analysis}
\begin{table*}
\centering
\begin{tabular}{ l l l l l l l l }
\hline 
\multicolumn{3}{c}{XRT} & & \multicolumn{3}{c}{BAT} \\
\cmidrule(lr){2-4}\cmidrule(lr){6-8}
{ObsID (*)} & {Obs. Date} & {Obs. Date} & {Exposure} & & {Obs. Date start} & {Obs. Date end} & {Exposure} \\
& (MJD) & (yyyy-mm-dd) & (ks) & &  (MJD) & (MJD) & (ks) \\
\hline
*20 & 57085.63121 & 2015-03-04 &  2.170 & & 57060 & 57100 & 657.0 \\
*27 & 57105.94254 & 2015-03-24 & 0.588 & & 57105.8 & 57112 & 131.8 \\
*28 & 57112.98183 & 2015-03-31 & 1.090 & & 57112 & 57114 & 35.3 \\
*29 & 57115.83525 & 2015-04-03 &  1.170 & & 57115 & 57117 & 47.2 \\
*30 & 57118.17901 & 2015-04-06 & 0.973 & & 57117 & 57119.5 & 46.4  \\
*31 & 57121.17200 & 2015-04-09 & 0.981 & & 57119.5 & 57122.5 & 59.5 \\
*32 & 57124.16573 & 2015-04-12 & 0.983 & & 57122.5 & 57125.5 & 71.5 \\
**01 & 57126.54499 & 2015-04-14 & 0.998 & & 57125.5 & 57130 & 99.7 \\
*34 & 57166.75520 & 2015-05-24 & 1.620 & & 57130 & 57165 & 714.2\\
\hline
\end{tabular}
\caption{Details of the Swift data used in the analysis of the 2015 outburst; *=000324360, **=000814510.}
\label{tab:obs}
\end{table*}
The XRT and BAT light curves and hardness ratios (HRs) collected during the bright 2015 outburst are shown in Fig. \ref{fig:swift}. \\
The flux increasing by an order of magnitude in the soft X-rays band around MJD 57115 and the corresponding decay in the XRT hardness ratio are signatures of the transition to the soft state \citep{Degenaar2015b}. A peculiar increase of the count rate in the hard X-rays, especially evident in the 15-20 keV range (Fig. \ref{fig:swift}), seems to anticipate the soft rise by $\sim$\,5 days. The nature of this hard rise will be discussed in the following. The XRT light curve profile shows a gap after the softening (see Fig. \ref{fig:swift}), but the flux decay can be followed using the \emph{MAXI} monitoring (see Fig. 1 in \cite{Degenaar2016}), which reveals the expected exponential decrease after the peak. \\ According to the two HR evolutions, we individuated three different phases (indicated with dashed lines in Fig. \ref{fig:swift}) until the soft X-rays peak: one long \emph{plateau} state, between $\sim$57060 and $\sim$57100 MJDs, characterized by a constant HR for both data-sets, a second shorter phase from $\sim$57106 MJD to $\sim$57112 MJD with two XRT observations, where an increase in the soft X-rays comes along with a simultaneous softening in the BAT data and finally a third phase, prior to the transition to the soft state, until 57114 MJD. According to this selection we performed a broad-band spectral analysis for each phase. Three time-averaged BAT spectra were therefore extracted, with number of channels chosen according to the statistics, i.e. 51 channels in the first two spectra and 18 in the third one. A single XRT observation representative of each phase was associated to each BAT spectrum. \\
For Phase 1, for which a sample of 14 XRT pointings were available with compatible flux and hardness ratio, we selected the observation placed in the middle of the time interval of interest (ObsID 00032436020). Between the two spectra in Phase 2, with consistent flux and HR, we chose the second, ObsID 00032436027, characterized by a better constrained HR. Finally, only one XRT observation was present within Phase 3 and it was paired with the corresponding BAT spectrum (Tab. \ref{tab:obs}). \\
The variable HR in the soft state did not allow for a similar grouping, therefore we proceeded differently. We extracted 5 subsequent BAT spectra, each one with a time-bin of $\approx$3 days and 12 channels, and we associated them to its corresponding XRT pointing. The sixth XRT spectrum is relative to an observation taken more than one month later: in order to collect as much statistics as possible and according to the constant BAT light curve in that period, we paired the spectrum with a six-weeks long BAT spectrum. In the final XRT observation, the low exposure, i.e. around 10 s, and the resulting low statistics did not enable for a reliable spectral analysis, therefore we decided to exclude the observation from our sample. The details about the observations used and the XRT and BAT pairings are reported in Table \ref{tab:obs}.\\

In every model used in this work, the component \textsc{tbabs} was included to take into account the photoelectric absorption due to neutral matter between the source and the observer, with photoelectric cross sections from \citet{Verner1996} and element abundances from \citet{Wilms2000}. Furthermore a \textsc{constant} component was included to serve as intercalibration constant. We also applied a systematic error of 2\%, according to the \swift\  calibration guidelines. In the following, the quoted errors are computed at 90\% confidence levels.\\
\subsection{Spectral modeling}
\label{ss:outburst}
Based on the \nustar\ spectral analysis reported by \cite{Ludlam2016}, RX1804 was in hard state during the BAT plateau, therefore we started testing a standard scenario for NS-LMXBs in hard state, i.e.  a multi-color disk black body plus a thermal Comptonization spectrum. We performed the spectral analysis of these three broad band XRT and BAT spectra in the energy range 1.0-120 keV. The Comptonization spectrum was described with the model \textsc{comptb} \citep{Farinelli2008}, used as a pure thermal one, i.e. with the logarithm of the illuminating factor $A$ and the bulk parameter $\delta$ frozen to respectively 8 and 0 \citep[as in][]{Cocchi2010}. The $\gamma$ parameter of the seed-photons distribution was fixed to the standard value of 3, corresponding to a pure black body distribution. The main parameters of this model are the corona temperature $kT_{\rm e}$, the temperature of the seed photons $kT_{\rm s,1}$ and $\alpha$, the energy index of the Comptonization spectrum. An estimate of the optical depth $\tau$ can be derived from the following equation \citep{Titarchuk1995}:
\begin{equation}
    \alpha=-\frac{3}{2}+\sqrt{\frac{9}{4}+\frac{\pi^2}{C_T}\frac{m_e c^2}{kT_e\left(\tau+\frac{2}{3}\right)^2}}
\end{equation}
 with the parameter $C_{\rm T}$ equal to 3 in spherical geometry. \\ The emission from the disk was modeled with the \textsc{Xspec} model \textsc{diskbb} \citep{Mitsuda1984}, whose normalization parameter $K$ is related to the apparent inner disk radius, given an assumed distance and inclination.
 
\begin{figure*}
\centering
\includegraphics[height=6 cm, width=8 cm]{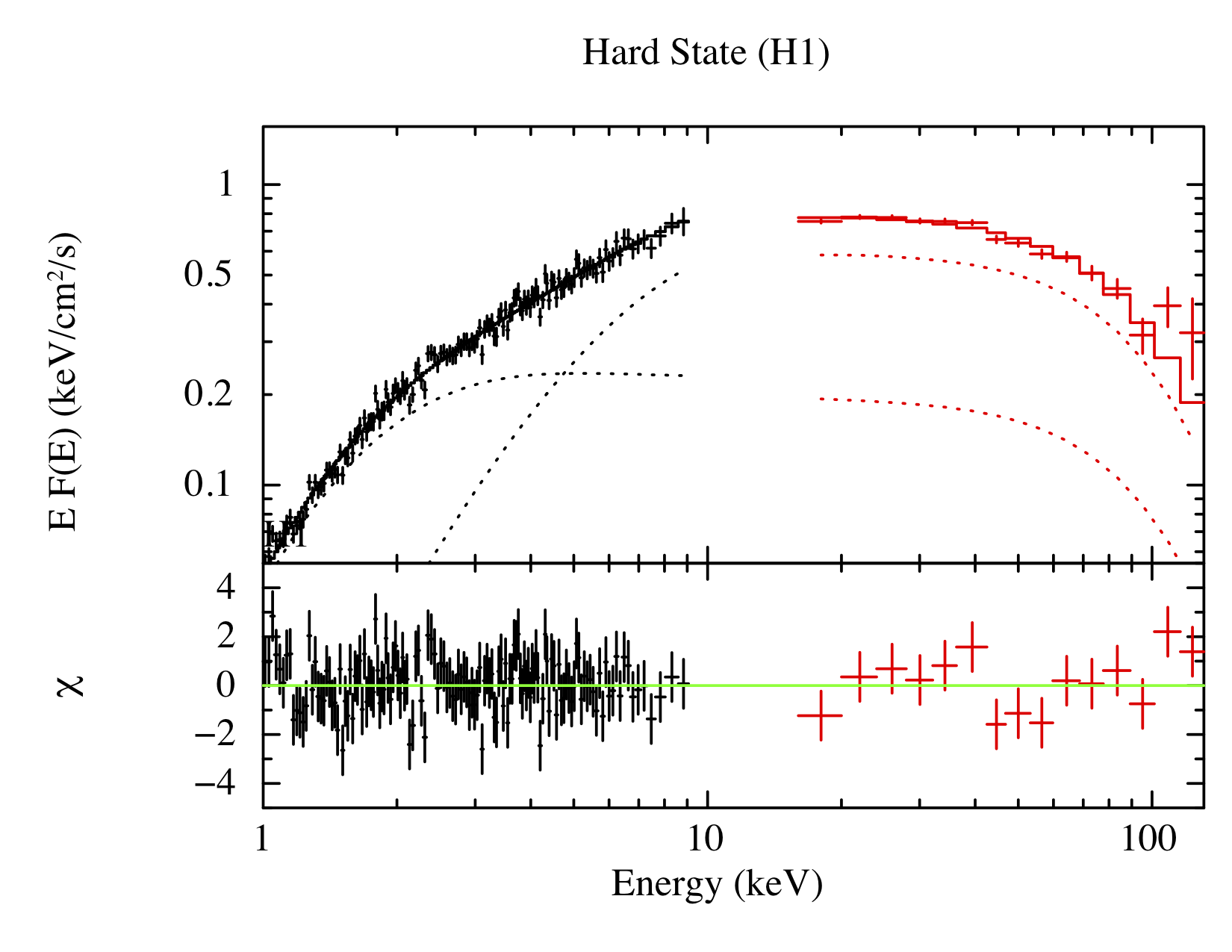}
\includegraphics[height=6 cm, width=8 cm]{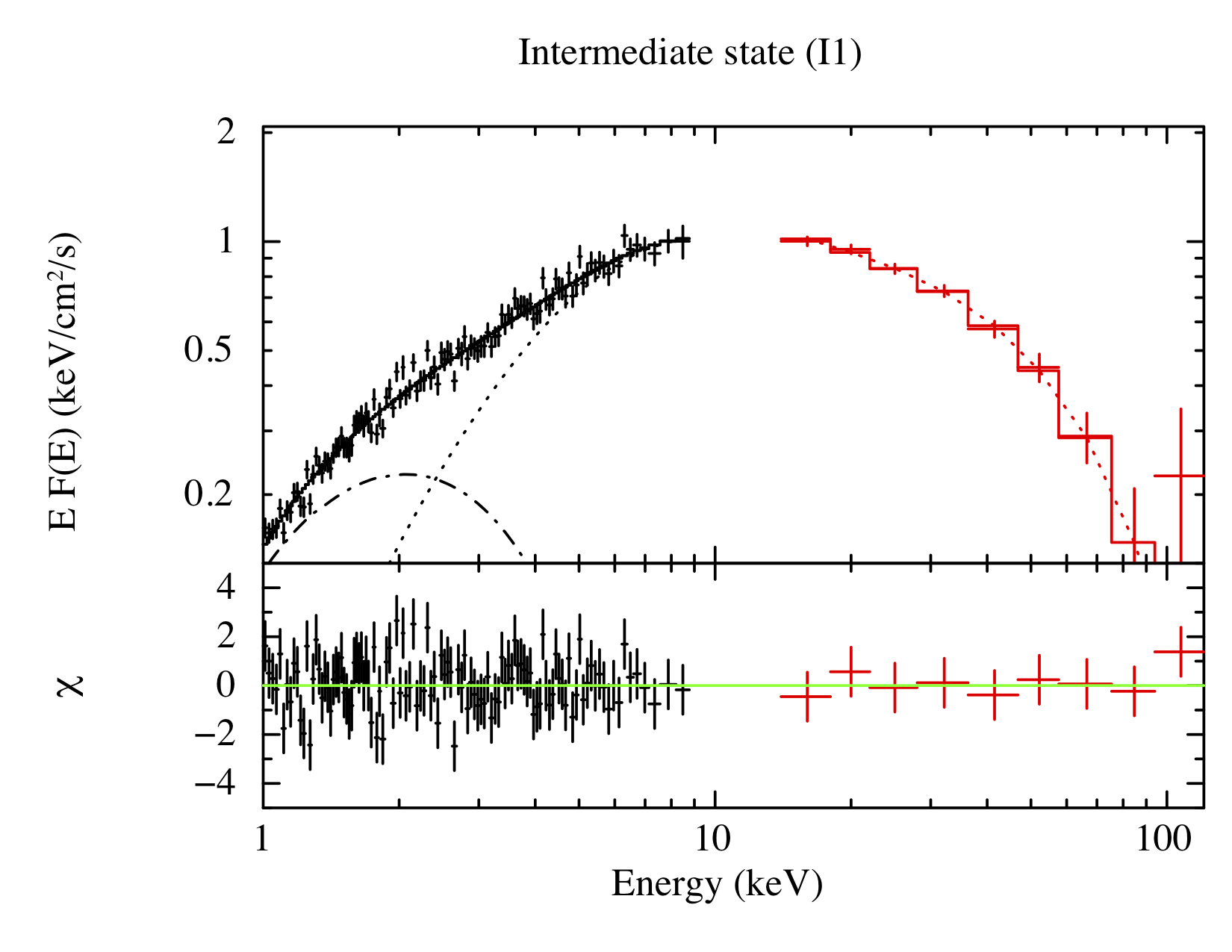}
\includegraphics[height=6 cm, width=8 cm]{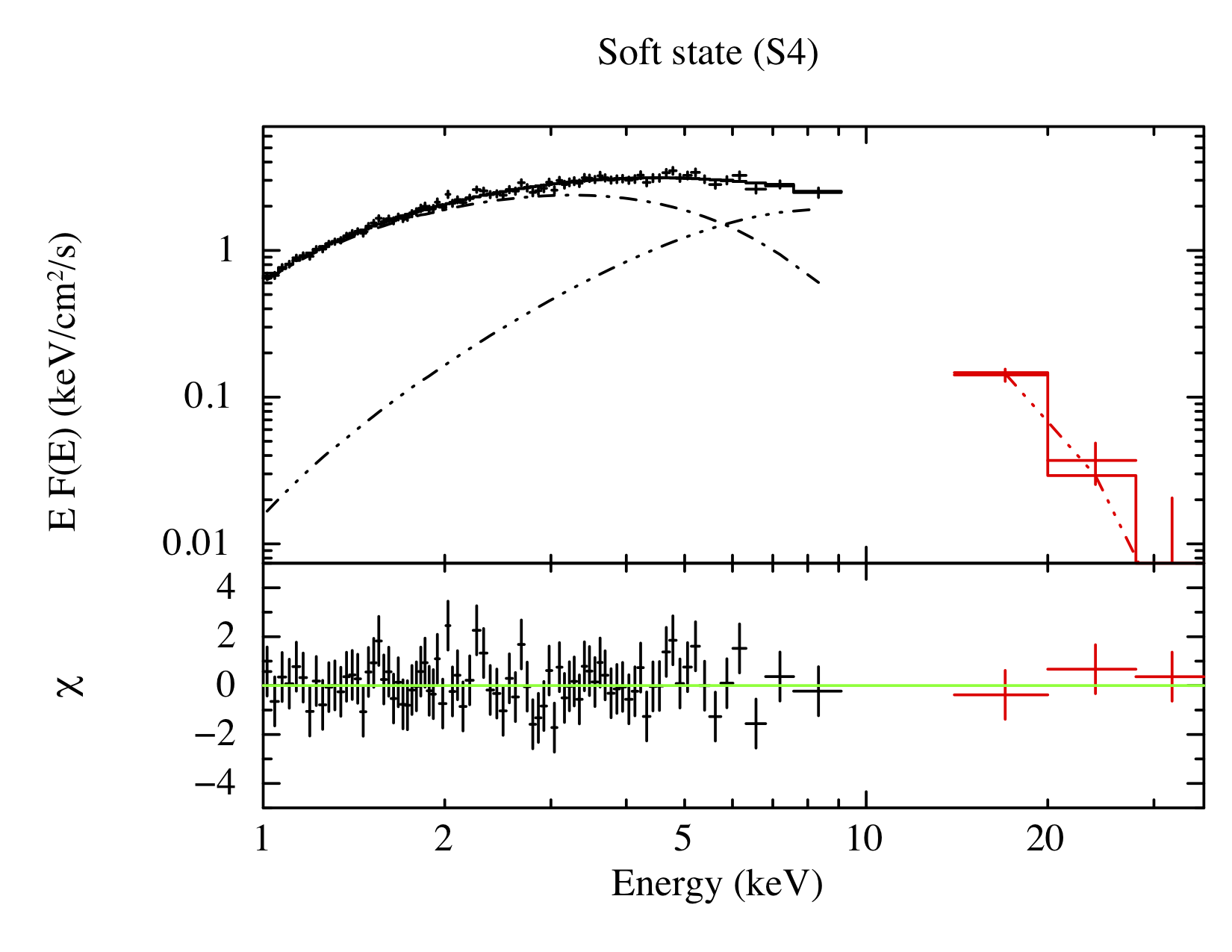}
\includegraphics[height=6 cm, width=8 cm]{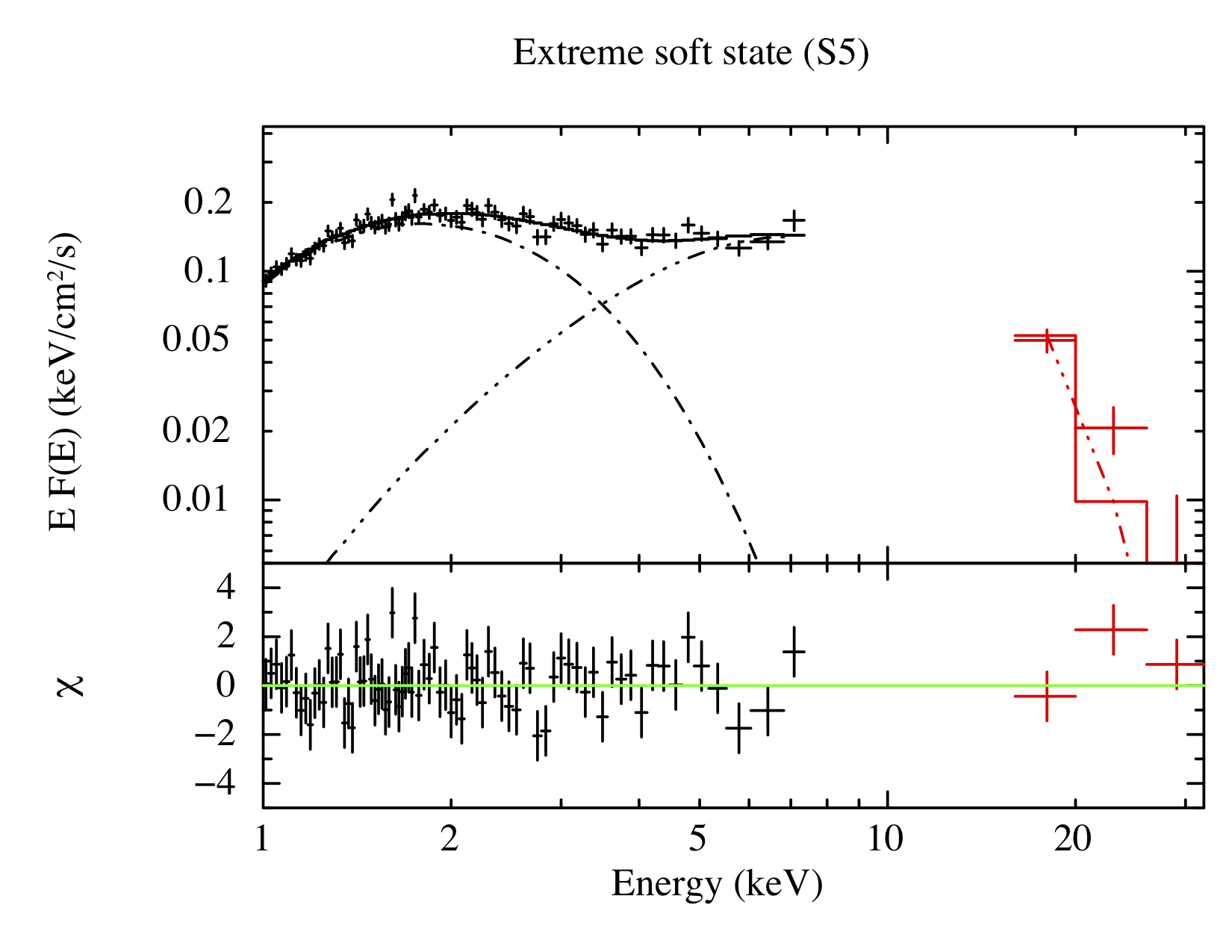}
\caption{XRT (\emph{black}) and BAT (\emph{red}) spectra of Obs. labeled as: H1 (\emph{top, left}), I1 (\emph{top, right}), S4 (\emph{bottom, left}) and S5 (\emph{bottom, right}) and relative residuals. In H1 we used the model \textsc{comptb}+\textsc{comptb}, 
in I1 we used \textsc{diskbb}+\textsc{comptb} while 
a \textsc{bbody}+\textsc{diskbb} model was employed for both S4 and S5. We adopted different linestyles to distinguish between the different components, in particular we used: dot for \textsc{comptb}, dash-dot for \textsc{diskbb} and dash-dot-dot-dot for \textsc{bbody}}.
\label{fig:swiftresiduals}
\end{figure*}

 Assuming a 10 kpc distance, the real inner radius of the disk was then found as a function of the inclination, using the correction factor $\kappa$=1.7 and a $\xi$ correction factor for the torque-free boundary condition of 0.4\footnote{This assumption is probably not valid for NS-LMXBs, therefore our values for the inner disk radius are likely underestimated.} \citep{Kubota1998, Gierlinski2002}.  \\ Although in  Obs. *20 and *27 the fits led to statistically good results, with a reduced $\chi^2_\nu$ (d.o.f.) of 1.04 (312) and 0.99 (151), respectively, the physical parameters we derived are quite odd: the seed-temperature of the Comptonization spectrum $kT_{\rm s,1}$ and the disk temperature $kT_{\rm disk}$, are indeed above $3$ and $1.7$ keV, respectively. The disk temperature seems hard to reconcile with a physically consistent scenario, since usually NS-LMXBs in hard state shows $kT_{\rm disk}$ much below 1 keV \citep[see, e.g., ][]{Dai2010,DiSalvo2015,Mazzola2019}. We checked if these results can be related to the lack of modeling of the reflection component, which was detected by \cite{Ludlam2016} in the same state. We applied the \textsc{reflect} convolution model \citep{Magdziarz1995} to the Comptonization spectrum in the three observations considered, fixing the abundances to the solar abundances and the inclination to 30$^\circ$ (according to the best-fit value found by \cite{Ludlam2016,Degenaar2016}).
 In both observations, the inclusion of this component was not statistically significant and did not modify the values of the two temperatures.  A more detailed study of the reflection component with the high statistics {\it NuSTAR} data is presented in Subsection \ref{sec:reflection}. Due to these results, we decided to rule out a \textsc{comptb}+\textsc{diskbb} model for the first two spectra in the hard state. \\
 We tried then to apply a double seed-photons Comptonization model, as suggested by several authors \citep[e.g.][]{Cocchi2011}. In this model, both populations of photons from the NS and the disk are seeds for the Comptonizing corona, characterized by a single electron temperature $kT_{\rm e}$ and a single optical depth. In order to describe this scenario, we summed in \textsc{Xspec} two different \textsc{comptb} components, with the cloud parameters, i.e. $kT_{\rm e}$ and the energy index of the Comptonization spectrum $\alpha$, tied between each other in the analogous models, following the prescription by \cite{Titarchuk2005}. On the other hand, we left the temperatures of the photon sources $kT_{\rm s,1}$, $kT_{\rm s,2}$ free and untied. The fit was unable to give any constraints on the $N_{\rm H}$ value, therefore this parameter was kept frozen to the best-fit value we found, i.e. 0.22$\times$10$^{22}$ cm$^{-2}$. \\
 On the contrary, Obs. *28 is not only statistically ($\chi^2_\nu$ (d.o.f.)=1.07(272)), but also physically well described by a \textsc{comptb}+\textsc{diskbb} model, with best-fit temperatures of $\sim$1.4 keV and $\sim$0.7 keV, respectively. For consistency, we applied the Comptonization model with double seed photons also to this spectrum. In this scenario, the seed photons temperatures obtained are physically consistent and compatible with NS (\emph{boundary layer}) and disk temperatures in hard state ($kT_{\rm s,1} \ \sim$1.4 and $kT_{\rm s,2} \ \sim$0.3 keV, respectively).
 However, while the values found for $kT_{\rm e}$ and $\tau$ in the first two spectra are typical of the hard states in NS-LMXBs (H1 and H2 in Tab. \ref{fig:swiftresiduals}), in the third spectrum a cooler $kT_{\rm e}$ has been found ($\sim$12 keV) with both models. A cooler corona implies the arising of a cooling source, e.g. the accretion disk, therefore a model including the direct emission from the disk, such as \textsc{comptb}+\textsc{diskbb}, could be a more physically acceptable scenario. Indeed, the observed increasing of the soft X-rays flux between the second and the third observation (see Fig. \ref{fig:swift}) is consistent with the arising of a contribution from a disk black body in the third spectrum. Based on these clues, we classified this spectrum as belonging to an intermediate state (I1 in Tab. \ref{tab:hard}). We show the energy spectra, the model components and the residuals for H1 and I1 in Fig. \ref{fig:swiftresiduals} (\emph{top}). \\

In the soft state spectra we ignored all the data beyond 40 keV, due to the lack of significant emission above that threshold. Furthermore, the poor spectral resolution of XRT, did not allow us to model the reflection features below 1 keV, apparent in the residuals in all the spectra in this state. Therefore we ignored all data below this threshold. \\
In order to be coherent with the analysis performed for I1, we applied a \textsc{comptb}+\textsc{diskbb} model to Obs. *29, finding an acceptable fit ($\chi^2_{\nu}$/d.o.f.=1.068(324)), but as long as we kept the (totally unconstrained) $\alpha$ parameter frozen to its best fit value of 1.2. The ongoing transition to the soft state is witnessed by the decreasing value of the corona temperature, about 3 keV and, by the increase of $kT_{\rm disk}$. Such a low value for $kT_{\rm e}$, with a higher $\tau$ compared to the previous spectrum (see Table \ref{tab:hard}), indicates the arising of a saturated Comptonization spectrum, typical of this state \citep[see, e.g. ][]{Barret2000,Gambino2019}. It is noteworthy that, in spite of the increasing $kT_{\rm disk}$, the disk is found at a higher $R_{\rm in}$, which is odd. This is probably due to the higher best-fit value of $N_{\rm H}$, which is known being strongly correlated with the \textsc{diskbb} normalization factor. Thus, we adopted the canonical soft state model, composed of a black body emission mimicking the saturated Comptonization (\textsc{bbody} in \textsc{xspec}) and the multicolor disk black body (\textsc{diskbb}). The fit with this model is acceptable as well ($\chi^2_{\nu}=$1.130(325)), and the physical parameters $kT_{\rm bb}$ and $kT_{\rm disk}$ are also compatible with those from the previous models. \\
In the remaining five observations, the latter model provides an acceptable fit and spectral parameters all in accordance with each other, with the exception of $kT_{\rm disk}$ in Obs. *34, which is significantly decreased in this last spectrum. On the contrary, a \textsc{diskbb}+\textsc{comptb} model does not provide a good fit in any of the last five spectra. So that, we have classified Obs. *29 as still an intermediate state, and labeled it as I2, and the observations from *30 to *34 as fully soft states (and labeled them from S1 to S5 in Tab. \ref{tab:hard}). In Fig. \ref{fig:swiftresiduals} (\emph{bottom}), the model components and residuals for S4 and S5 are shown.

\begin{table*}
\rotatebox{270}{
\begin{tabular}{l l l l l l l l l l l l l}
\hline
\hline
{\bf Model} & \multicolumn{2}{c}{\textbf{Parameters}} & & \multicolumn{9}{c}{\textbf{Spectra}} \\
\cmidrule(lr){4-13} 
& & & & H1 & H2 & I1 & I2 & S1 & S2 & S3 & S4 & S5 \\
\hline
\multirow{7}{*}{\textsc{comptb}+\textsc{comptb}} & $N_{\rm H}$ & 10$^{22}$ cm$^{-2}$ & & (0.22) & (0.22) & (0.22) & - & - & - & - & - & - \\ 
& $kT_{\rm e}$ & keV & & 20.0$^{+3.0}_{-1.5}$ & 17.6$^{+2.0}_{-1.7}$ & 11.8$^{+2.0}_{-1.5}$ & - & - & - & - & - & - \\
& $kT_{\rm s,1}$ & keV & & 2.1$^{+1.2}_{-0.5}$ & 1.5$^{+0.3}_{-0.2}$ & 1.20$^{+0.11}_{-0.10}$ & - & - & - & - & - &  - \\ 
& $kT_{\rm s,2}$ & keV & & 0.52$^{+0.07}_{-0.06}$ & 0.36$\pm$0.04 & 0.24$\pm$0.02 & - & - & - & - & - \\
& $\alpha$ & & & 1.04$^{+0.11}_{-0.05}$ & 1.20$\pm$0.06 & 1.15$\pm$0.13 & - & - & - & - & - & - \\
& $\tau$ & & & 4.00$\pm$0.06 & 3.65$\pm$0.14 & 4.80$\pm$0.40 & - & - & - & - & - & - \\
& {$\chi^2_\nu$} & (d.o.f.) & & 1.028 (312) & 0.929(151) & 1.080 (242) & - & - & - & - & - & - \\
\cmidrule{2-13}
\multirow{8}{*}{\textsc{comptb}+\textsc{diskbb}} &  $N_{\rm H}$ & 10$^{22}$ cm$^{-2}$ & & - & - & (0.22) & 0.36$^{+0.04}_{-0.03}$& - & - & - & - & - \\
& $kT_{\rm e}$ & keV & & - & - & 13.0$^{+3.0}_{-2.0}$ & <3.4 & - & - & - & - & -\\
& $kT_{\rm s}$ & keV & & - & - & 1.36$^{+0.19}_{-0.14}$ & 1.2$^{+3.0}_{-0.4}$ & - & - & - & - & - \\
& $kT_{\rm disk}$ & keV & &  - & - & 0.70$^{+0.09}_{-0.08}$ & 1.0$^{+0.4}_{-0.3}$ & - & - & - & - & - \\
& $R_{\rm in}\cos{i}$  & km & & - & - & 13.4$^{+0.6}_{-0.5}$ & 23.9$^{+3.0}_{-1.6}$ & - & - & - & - & -\\
& $\alpha$ & & &  - & - & 1.24$\pm$0.13 & (1.2) & - & - & - & - & - \\
& $\tau$ & & &  - & - & 4.4$\pm$0.4 & 10.7$\pm$0.2 & - & - & - & - & - \\
& {$\chi^2_\nu$} & (d.o.f.) & & - & - & 1.074(242) & 1.068(324) & - & - & - & - & - \\
\cmidrule{2-13}
\multirow{5}{*}{\textsc{bbody}+\textsc{diskbb}} & $N_{\rm H}$ & 10$^{22}$ cm$^{-2}$ & & - & - & - & 0.25$\pm$0.02 & 0.39$\pm$0.03 & 0.46$\pm$0.04 & 0.27$\pm$0.02 & 0.22$\pm$0.02 & 0.29$\pm$0.08  \\
& $kT_{\rm bb}$ & keV & & - & - & - & 2.0$^{+0.4}_{-0.3}$& 1.9$\pm$0.2 & 2.1$^{+0.5}_{-0.3}$ & 2.4$^{+0.7}_{-0.4}$ & 2.3$\pm$0.4 & 1.8$^{+0.3}_{-0.2}$ \\
& $kT_{\rm disk}$ & keV & & - & - & - & 1.18$\pm$0.15 & 1.08$^{+0.12}_{-0.11}$ & 1.20$^{\pm}$0.17 & 1.27$^{+0.20}_{-0.17}$ & 1.39$^{+0.16}_{-0.13}$ & 0.69$\pm$0.04  \\
& $R_{\rm in}\cos{i}$ & km & & - & - & - & 17.8$^{+0.9}_{-0.7}$ & 21.0$^{+1.0}_{-0.8}$ & 17.7$\pm$0.6 &  14.0$^{+0.9}_{-0.6}$ & 13.0$^{+0.6}_{-0.4}$ & 15.7$^{+0.8}_{-0.6}$ \\
& {$\chi^2_\nu$} & (d.o.f.) & & - & - & - & 1.13(325) & 1.15(287) & 1.13(297) & 1.14(250) & 0.98(267) & 1.13(180) \\
\hline
\multirow{2}{*}{\textbf{Flux} $[\times 10^{-9}]$} & Bol & \ergcm & & 3.7 & 3.9 & 4.7 & 16.6 & 17.0 & 18.6 & 13.3 & 12.0 & 1.8\\
& Disk & \ergcm & & - & - & 0.7 & 7.9 & 9.2 & 10.2 & 7.6 & 8.0 & 1.2 \\
\hline
\end{tabular}
}
\caption{Fit results of the combined XRT and BAT spectra collected during the 2015 outburst. Quoted errors reflect 90\% confidence level. The parameters which were kept frozen during the fits are reported between round parentheses. The reported flux values have associated errors of 10\%.}
\label{tab:hard}
\end{table*}

\section{Studying the reflection component with \nustar}
\label{sec:reflection}
The spectral analysis performed in Subsection \ref{ss:outburst} had the goal to study the spectral evolution of the source during the outburst. Since the poor statistics of our \swift\ spectra did not allow us to probe the presence of the reflection, we combined our broad band XRT and BAT spectra with three \nustar\ observations (see Table \ref{tab:nustar}).    
\begin{figure}
\centering
\includegraphics[height=6 cm, width=8 cm]{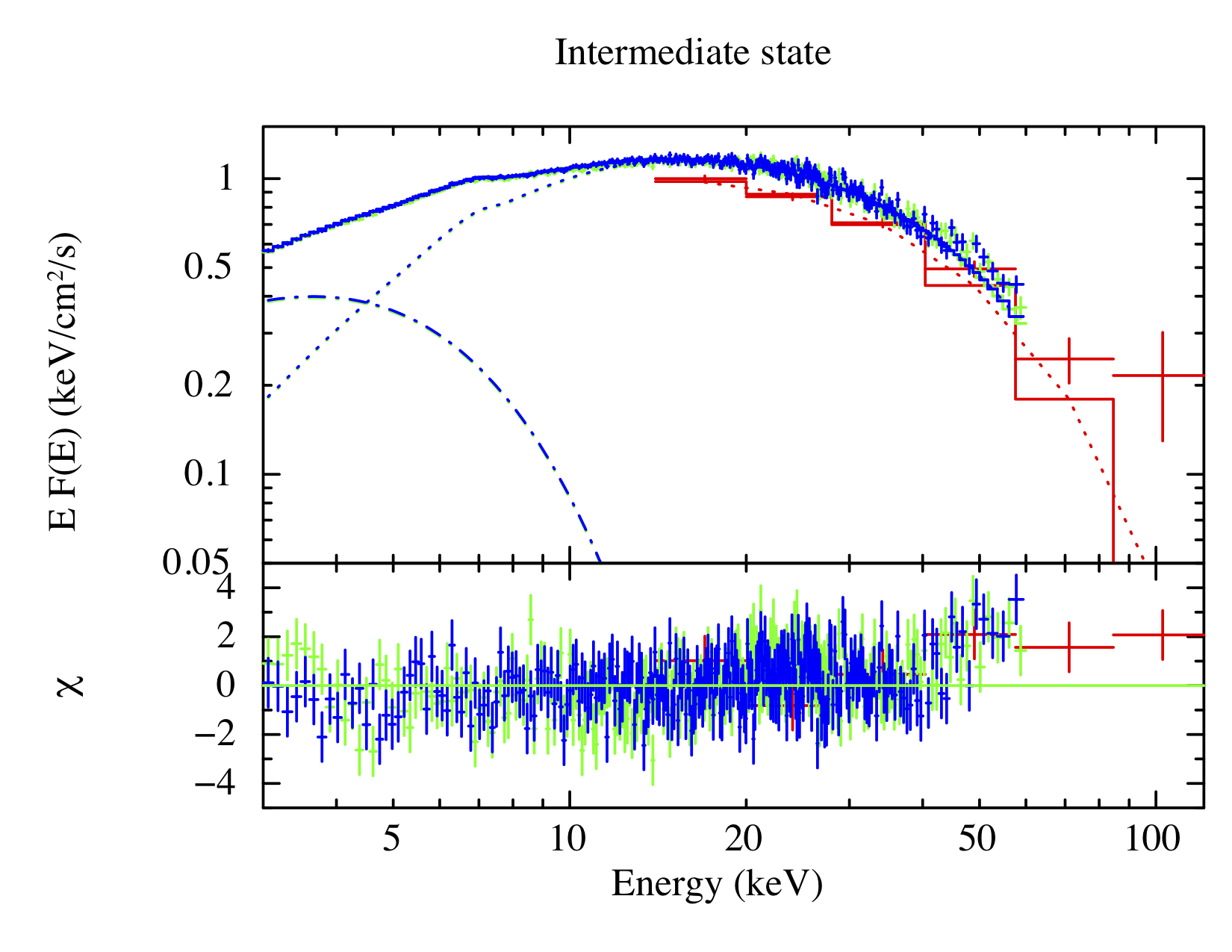}
\caption{BAT+\nustar\ energy spectrum in the intermediate state fitted with a Comptonization (dot line) plus a disk model (dash-dot line). Reflection component is also included.  Residuals at high energy (above 40 keV) are significant.}
\label{fig:residuals}
\end{figure}

\begin{figure}
\centering
\includegraphics[height=6 cm, width=8 cm]{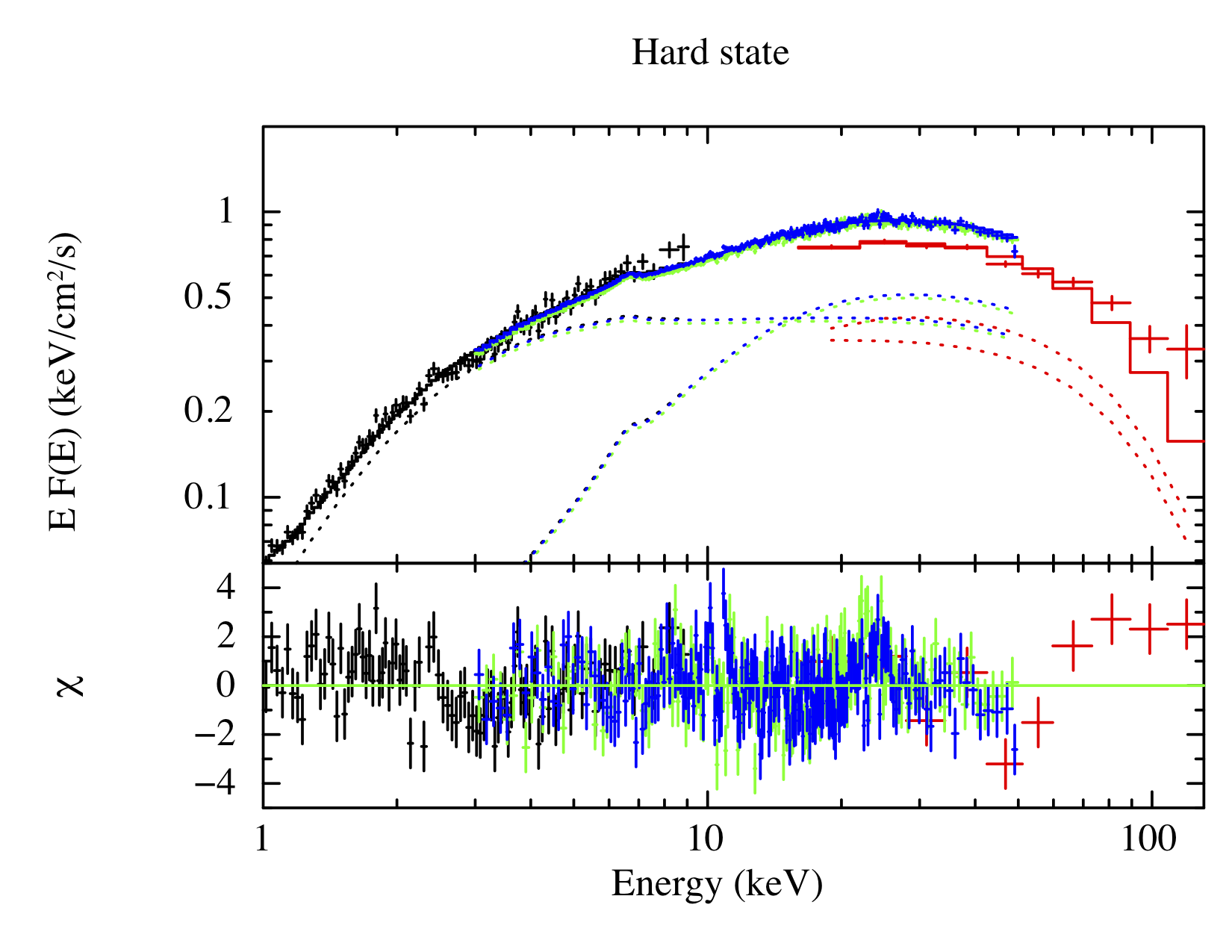}
\includegraphics[height=6 cm, width=8 cm]{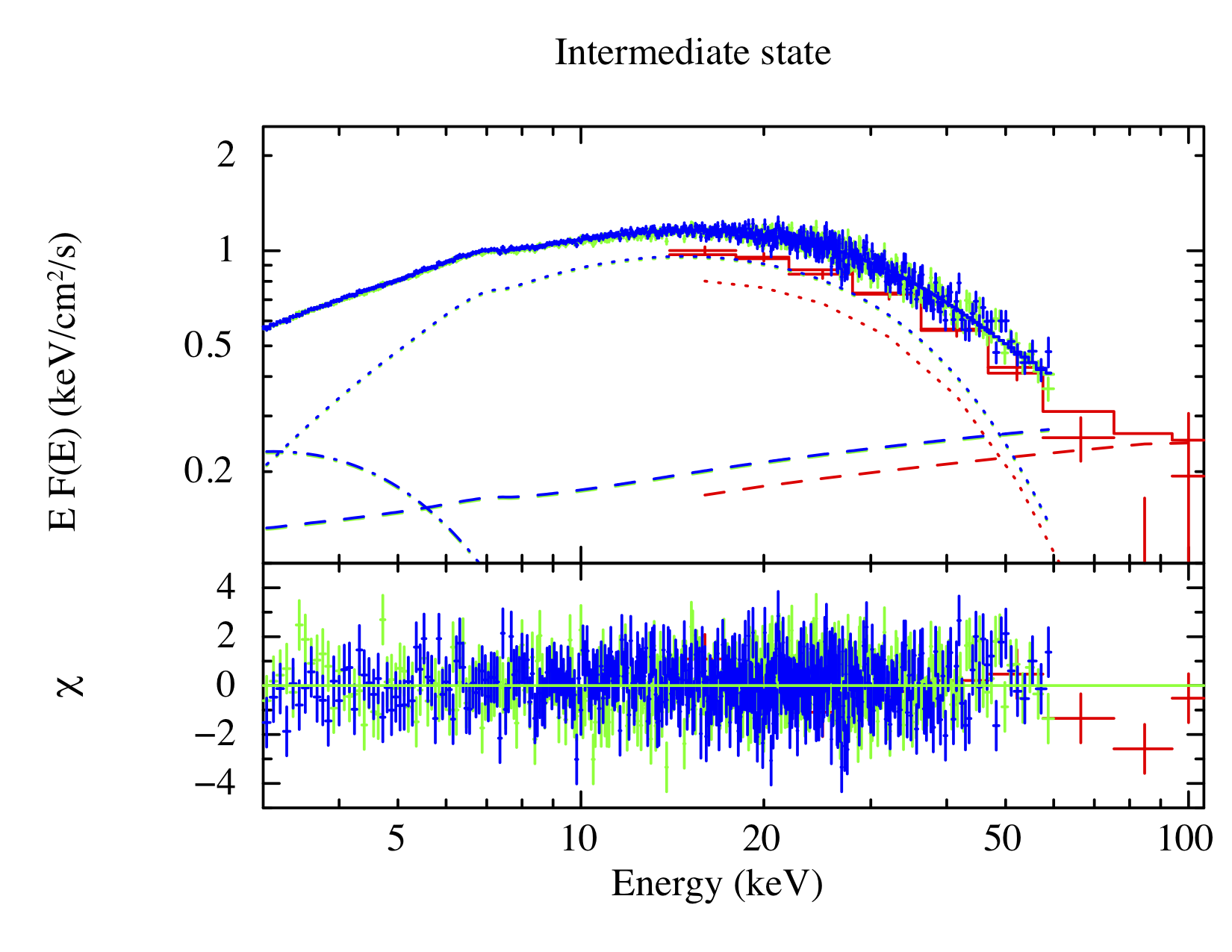}
\includegraphics[height=6 cm, width=8 cm]{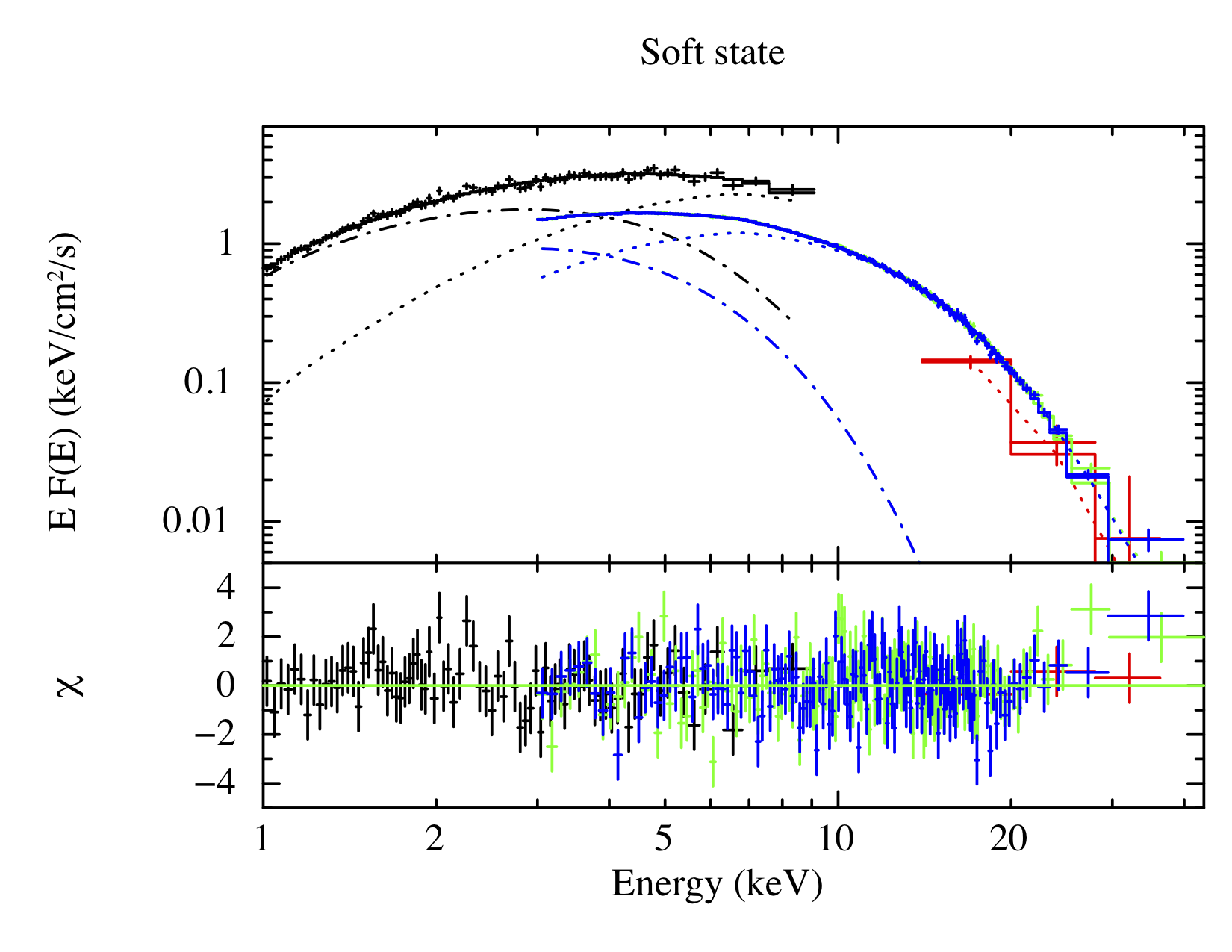}
\caption{Energy spectra and residuals of RX1804 in hard (\emph{top}), intermediate (\emph{middle}) and soft (\emph{bottom}) state including XRT + BAT + \nustar\ for the \emph{top} and \emph{bottom} spectra and only BAT + \nustar\ for the \emph{middle} one. We adopted different linestyles to distinguish between the different components, in particular we used: dot for \textsc{comptb}, dash-dot for \textsc{diskbb} and dash for \textsc{powerlaw}.}
\label{fig:nustar}
\end{figure}

We first performed a preliminary analysis of the iron line using only \nustar\ data. We modeled the continuum with the same models used in this work for the XRT and BAT in hard, intermediate and soft states, while a simple \textsc{gaussian} component was included to describe the iron line. In all the three \nustar\ spectra, the line was found at an energy comprised between 6.5 and 6.7 keV, indicating a relatively high ionisation for the accreting material in all states. Furthermore, the equivalent width of the line was found around 50-70 eV, 75-110 eV and 40-100 eV, respectively. This points out a higher contribution of reflection in the intermediate state with respect to the hard state, while a similar consideration for the soft state is not possible, since the high uncertainties. We then associated each of the \nustar\ observations to the quasi-simultaneous \swift\ spectrum in Table \ref{tab:obs}. In particular we coupled \nustar\ Observation 80001040002 with H1, Obs. 80001040004 with I1 and Obs. 90101003002 with S4. \\
We used the convolution model \textsc{rfxconv} \citep{Done2006,Kolehmainen2011}, which is a self-consistent model including the effects of the absorption edges and the emission lines expected for reflection from accretion disks in X-ray binaries. We also applied the convolution model \textsc{rdblur} to smear the spectrum simulating the relativistic effects due to the NS gravitational field and the dynamics of the accretion disk. We applied \textsc{rdblur}$\times$\textsc{rfxconv} to both the Comptonization components\footnote{We tied together all the parameters in the reflection components applied to the two Comptonization spectra, as again we are considering in this scenario a single hot electron plasma.} in the hard state, and to the single \textsc{comptb} component in the other two states. \\
In order to reduce the number of degeneracies, we fixed the outer radius of the disk to 1000 R$_{\rm G}$ and the iron abundance to the solar abundance. The emissivity index $Betor$\footnote{the exponent of the distance $r$ to the center of the disk if we scale emissivity as $r^{Betor}$} was not well constrained by the fit, i.e. it was found in a wide range between -2 and -3.5. It was then fixed to the reasonable value of $-2.7$. We left free, instead, the ionization parameter and the inclination. However, due to the strong correlation between the inclination and the inner disk radius we were not able to find constraints on both the parameters. We considered fixing the inclination a viable way to obtain constraints on the inner disk radius. Moreover, due to the ignorance about the real inclination of the system, we attempted to calculate $R_{\rm in}$ considering a high and a low inclination cases, corresponding respectively to 60$^\circ$ (since no dips and/or eclipses have been ever found) and 20$^\circ$.  
The fit to the broad band XRT, \nustar\ and BAT spectrum with the reflection-inclusive best-fit models individuated in Subsection \ref{ss:outburst}, reveals unmodeled residuals above $\sim$40 keV and a general bad accordance with the data in the intermediate state (I1). Since the spectra in this state were taken during the rapid hard-to-soft state transition, 
we checked whether XRT and \nustar\ observations, performed only $\sim$17 hours apart, could have incompatible 
spectral shapes fitting them separately with a \textsc{powerlaw} model and allowing the $\Gamma$ index 
to vary between the two spectra. The best-fit indices are not compatible within the errors, pointing out that the RX1804 spectrum was varied, though XRT and \nustar\ observations should not be fitted together. We then paired \nustar\ with BAT, 
since the spectrum of the latter was averaged over a large time interval, compatible with the exposure time of the \nustar\ spectrum. Fitting \nustar\ and BAT with the same model used before gives an acceptable fit, but in both spectra we saw clearly residuals at high energies, which reveal the presence of a hard tail (Fig. \ref{fig:residuals}). In order to improve the fit, we included a \textsc{powerlaw} component, convolved with \textsc{rdblur}$\times$\textsc{rfxconv}. The new component improves significantly the fit, i.e. \emph{f-test} probability of improvement by chance of $\sim$1e-11, confirming the arising of a hard tail in the intermediate state, with a flux of $\sim$4$\times$10$^{-10}$ \ergcm\ in the 30-100 keV energy band, a factor of 2 lower of the total flux in the same band. \\ We added a power-law component also to the models used for the analysis of the hard and soft states, in order to check for the presence of any hard tail in these states. In hard state, the introduction of a reflected power-law component improves significantly the fit (\emph{f-test} probability of $\sim$1.0e-14), but at the cost of making the other physical parameters of the fit, especially $kT_{\rm e}$, undetermined. A similar situation concerns the soft state spectra, where in spite of a slight improvement in the fit (\emph{f-test} probability of $\sim$1.0e-07), the $\Gamma$ index of the reflected power-law was completely unconstrained by the fit. We therefore conclude that, while a hard tail contribution is likely to be taken into account at every spectral state of the source, we have the statistics to probe it properly only in the intermediate state. 
The results of the fits of the three spectra are reported in Table \ref{tab:refl}, while the energy spectra plus residuals are shown in Figure \ref{fig:models} \\
A comparison between these results and the results in Table \ref{tab:hard} reveals how the inclusion of the reflection component does not change significantly the values for the tempeatures $kT_{\rm s,1}$, $kT_{\rm s,2}$, $kT_{\rm e}$ and $kT_{\rm disk}$, except for H1, where the value of $kT_{\rm s,1}$ increases significantly. This might be due to a hidden contribution of an unmodeled power-law component at high energies, which drives the free parameters, and in particular $kT_{\rm s,1}$, to a physically odd minimum. In the three spectra the reflection fraction was always found lower than 20\%, indicating a consistent but not extremely strong reflection component. Ionization was found relatively high, i.e. $\log \xi \approx$2-3, in all the spectra, which is in accordance with the simple line analysis performed with a \textsc{gaussian} model in the \nustar-only spectra. This value is also in agreement with those found by \cite{Ludlam2016} and \cite{Degenaar2016} in hard and soft state, respectively. The wide uncertainties in the inner radius values in the three states do not allow us to strongly identify constraints for $R_{\rm in}$, which was found to be <90 $R_{\rm G}$ in H1, between 16 and 40 $R_{\rm G}$ in I1 and >13 $R_{\rm G}$ in S4.\\ The bolometric unabsorbed fluxes found from the broad band analysis are compatible within the errors with the fluxes found in the \swift\ -only analysis, i.e. $F_{\rm 0.1-150 keV}\sim$4.0$\times10^{-9}$ \ergcm for H1, $F_{\rm 0.1-150 keV}\sim$5.0$\times10^{-9}$ \ergcm for I1 and $F_{\rm 0.1-50 keV}\sim$11.0$\times10^{-9}$ \ergcm for S4. 

\begin{table*}
\centering
\begin{tabular}{ l l l l l l l l l l l l }
\hline
\hline
 \multicolumn{12}{c}{Model: \textsc{rdblur}$\times$\textsc{rfxconv}$\times$(\textsc{comptb}+\textsc{comptb})} \\
 \multicolumn{12}{c}{\bf H1}\\
\cmidrule{1-12}
$N_{\rm H}$ & $kT_{\rm s,1}$ & $kT_{\rm s,2}$ & $kT_{\rm e}$ & $f_{\rm refl}$ & $R_{\rm in}$ & i & $\alpha$ & $\tau$ & $\log{\xi}$ & & $\chi^2_{\nu}$\\
$\times$10$^{22}$ \ cm$^{-2}$ & keV & keV & keV & & $R_{\rm G}$ & $^\circ$ & & & & & (d.o.f.)\\
\hline
(0.22) & 3.19$^{+0.14}_{-0.15}$ & 0.75$\pm$0.02 & 17.3$^{+0.7}_{-0.6}$ & 0.16$^{+0.05}_{-0.04}$ & 30$^{+60}_{-20}$ & (60) & 0.97$\pm$0.02 & 4.39$\pm$0.07 & 2.37$^{+0.05}_{-0.03}$ & & 1.16(1462) \T\\
(0.22) & 3.64$^{+0.14}_{-0.11}$ & 0.83$^{+0.03}_{-0.02}$ & 18.5$^{+0.9}_{-0.7}$ & 0.034$\pm$0.003 & <21 & (20) & 1.01$\pm$0.03 & 4.04$\pm$0.09 & 2.84$^{+0.10}_{-0.05}$ & & 1.14(1462) \T \B\\
\hline
\multicolumn{12}{c}{Model: \textsc{rdblur}$\times$\textsc{rfxconv}$\times$(\textsc{comptb}+\textsc{powerlaw})+\textsc{diskbb}} \\
\multicolumn{12}{c}{\bf I1}\\
\cmidrule{1-12}
$N_{\rm H}$ & $kT_{\rm s,1}$ & $kT_{\rm disk}$ & $kT_{\rm e}$ & $f_{\rm refl}$ & $R_{\rm in}$ & i & $\alpha$ & $\tau$ & $\log{\xi}$ & $\Gamma$ & $\chi^2_{\nu}$ \\
$\times$10$^{22}$ \ cm$^{-2}$ & keV & keV & keV & & R$_{\rm G}$ & $^\circ$ & & & & & (d.o.f.) \\
\cmidrule{1-12}
(0.22) & 1.7$\pm$0.2 & 1.30$^{+0.10}_{-0.16}$ & 7.5$^{+0.5}_{-0.4}$ & 0.13$\pm$0.3 & 27$^{+12}_{-8}$ & (60) & 0.96$^{+0.05}_{-0.02}$ & 7.0$\pm$0.2 & 2.40$^{+0.04}_{-0.03}$ & 1.80$^{+0.11}_{-0.18}$ & 0.99(1561) \T\\
(0.22) & 1.59$^{+0.10}_{-0.08}$ & 1.24$^{+0.05}_{-0.02}$ & 7.4$^{+0.3}_{-0.2}$ & 0.052$^{+0.006}_{-0.005}$ & 25$^{+11}_{-9}$ & (20) & 0.91$^{+0.03}_{-0.02}$ & 7.5$\pm$1.2 & (2.4) & (1.77) & 0.98(1561) \T \B\\
\hline
\multicolumn{12}{c}{Model: \textsc{rdblur}$\times$\textsc{rfxconv}$\times$\textsc{comptb}+\textsc{diskbb}} \\
\multicolumn{12}{c}{\bf S4}\\
\cmidrule{1-12}
$N_{\rm H}$ & $kT_{\rm s,1}$ & $kT_{\rm disk}$ & $kT_{\rm e}$ & $f_{\rm refl}$ & $R_{\rm in}$ & i & $\alpha$ & $\tau$ & $\log{\xi}$ & & $\chi^2_{\nu}$ \\
$\times$10$^{22}$ \ cm$^{-2}$ & keV & keV & keV & & R$_{\rm G}$ & $^\circ$ & & & & & (d.o.f.) \\
\cmidrule{1-12}
0.31$\pm$0.03 & 1.31$\pm$0.03 & (1.16) & 2.41$^{+0.04}_{-0.03}$ & 0.07$^{+0.10}_{-0.09}$ & >25 & (60)  & (1.2) & 11.1$\pm$0.1 & 3.0$^{+0.1}_{-0.3}$ & & 1.08(1074) \T\\
0.28$\pm$0.02 & 1.03$^{+0.07}_{-0.05}$ & (0.90) & 2.49$^{+0.06}_{-0.04}$ & 0.09$^{+0.05}_{-0.02}$ & 23$^{+65}_{-13}$ & (20) & (1.2) & 10.9$\pm$0.01 & 3.3$\pm$0.1 & & 1.07(1076) \T \B\\
\hline
\hline
\end{tabular}
\caption{Fit results from the combined XRT, BAT and \nustar\ spectra. Quoted errors reflect 90\% confidence level. Fixed parameters are in round brackets. See text for the parameters definition.}
\label{tab:refl}
\end{table*}

\section{Type-I X-ray Bursts data analysis}
\label{sec:burst}

\begin{figure*}
\centering
\includegraphics[height=6.5 cm, width=9 cm]{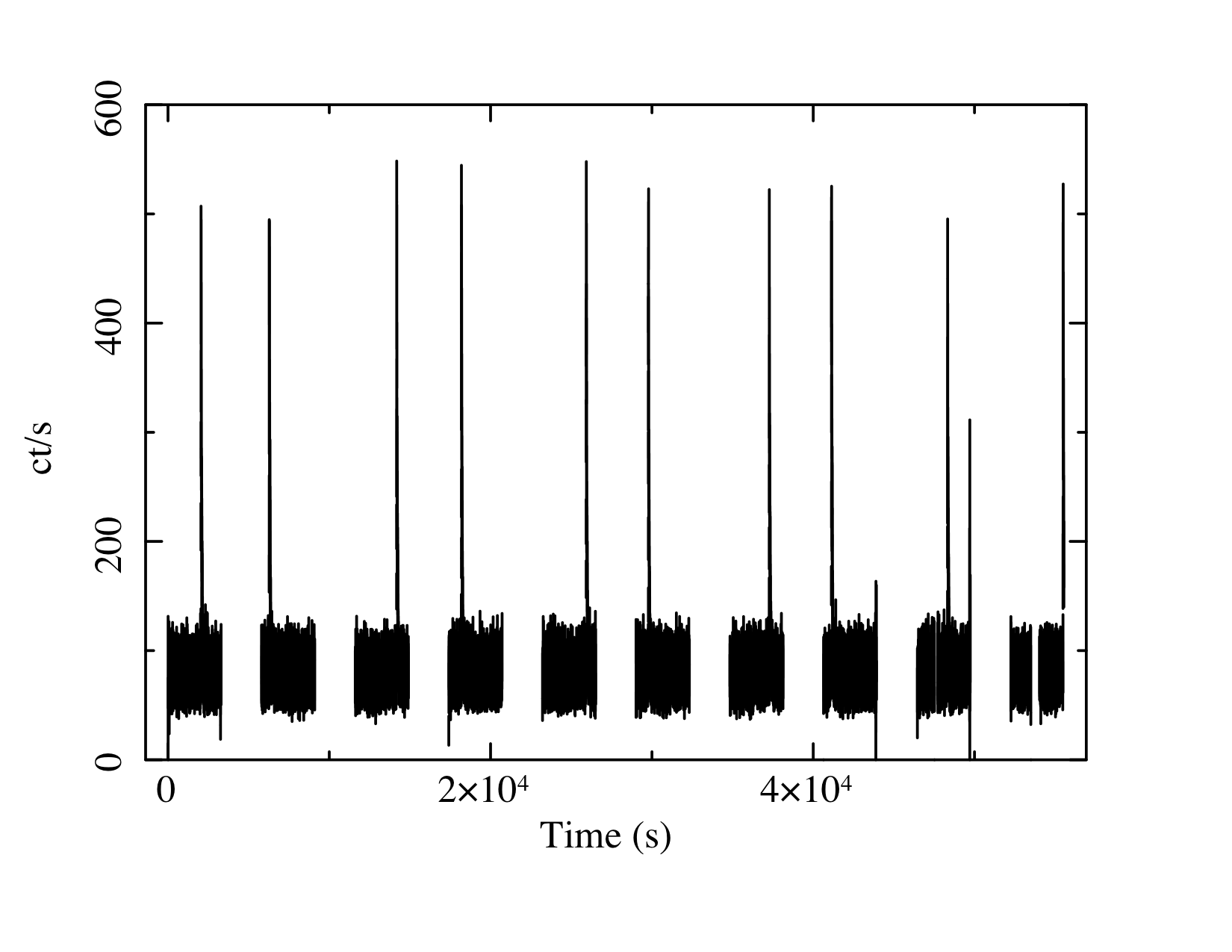}
\includegraphics[height=6.5 cm, width=8.5 cm]{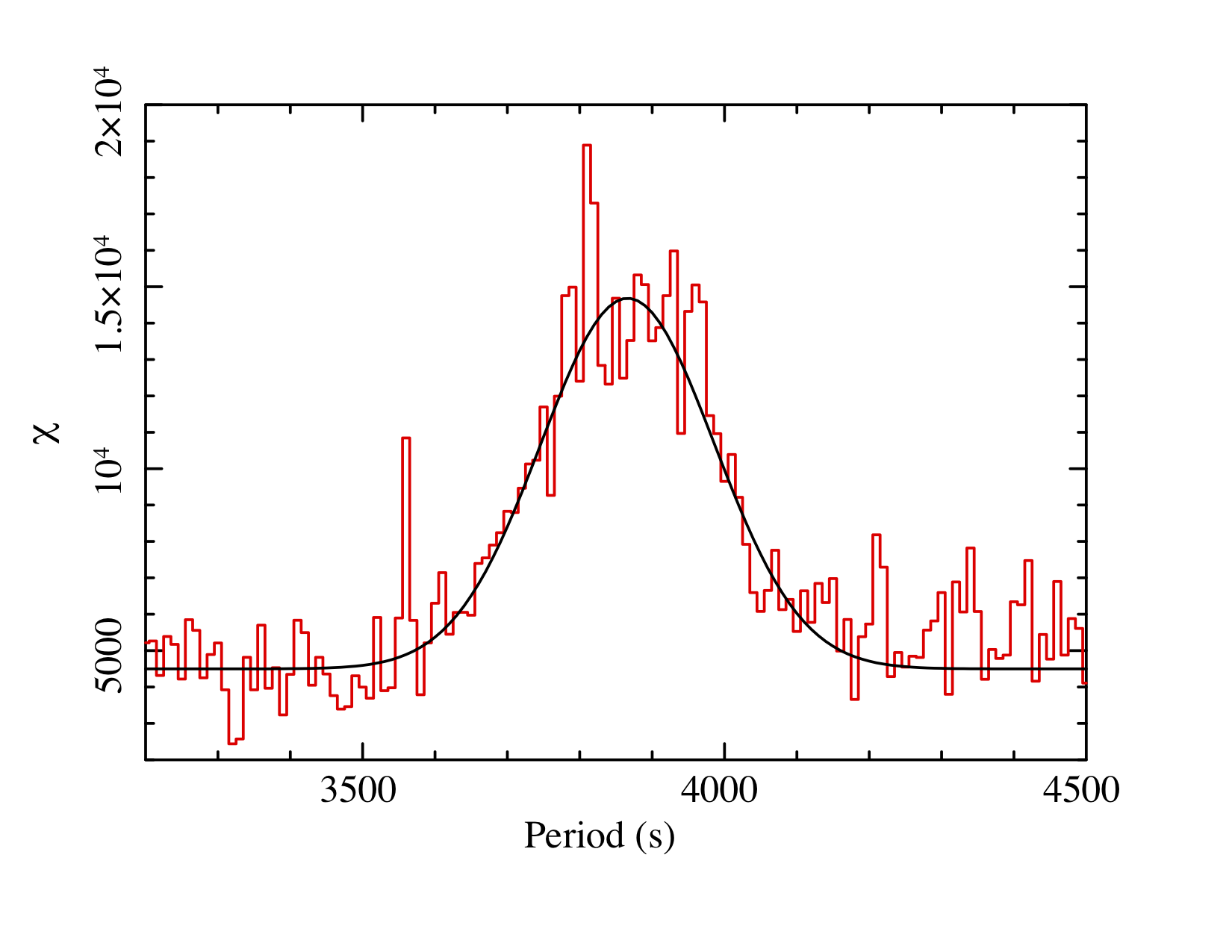}
\caption{Light curve of the ten type-I bursts observed by \nustar\ in 2015 during the intermediate state (\emph{left}) and folding of the light curve, fitted with a gaussian model (\emph{right}). The zero in the light curve corresponds to the Modified Julian Date (MJD) 57113.6798.}
\label{fig:clock}
\end{figure*}

\begin{figure*}
\centering
\includegraphics[height=6 cm, width=8.5 cm]{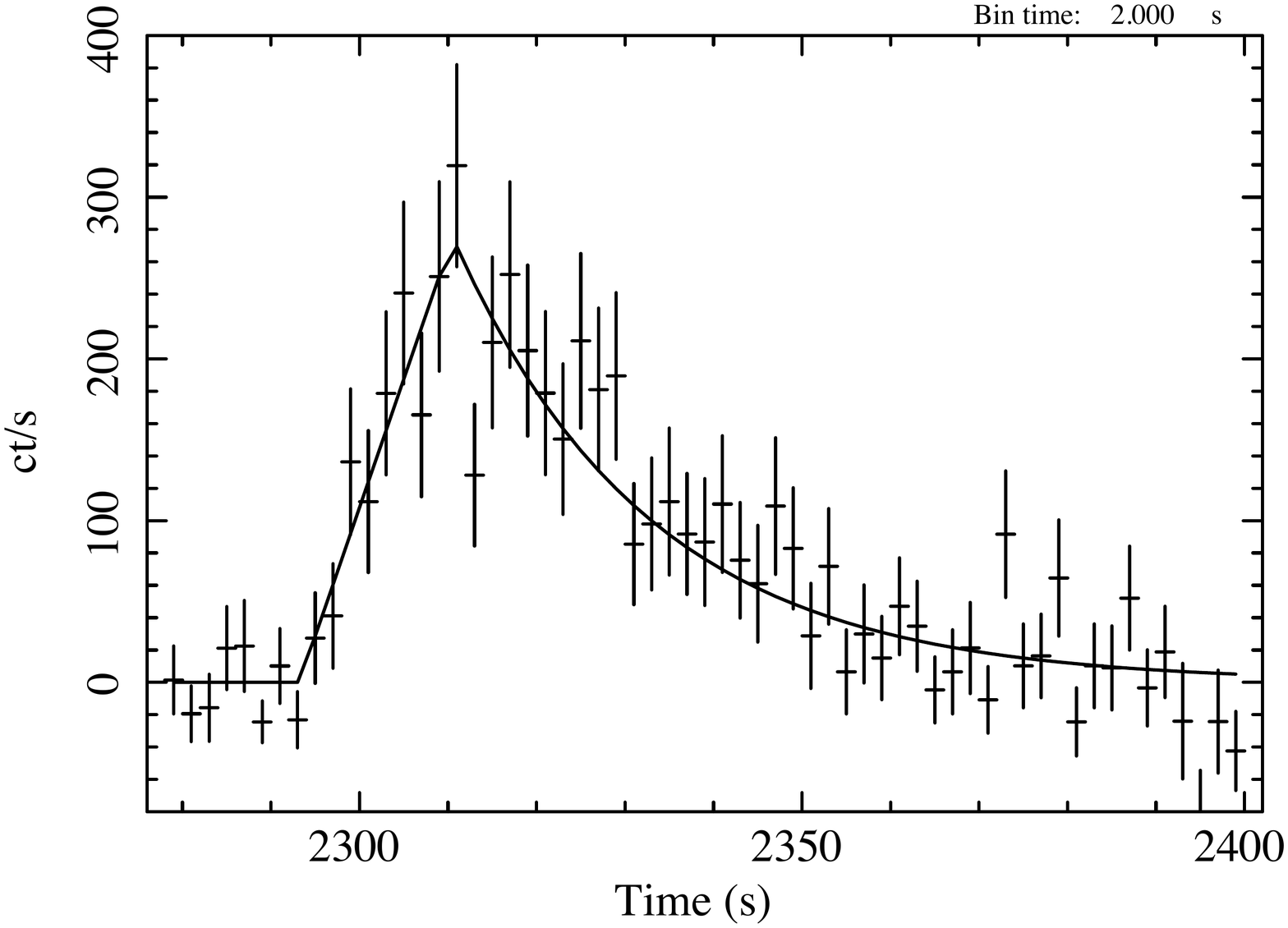}
\includegraphics[height=6.5 cm, width=8 cm]{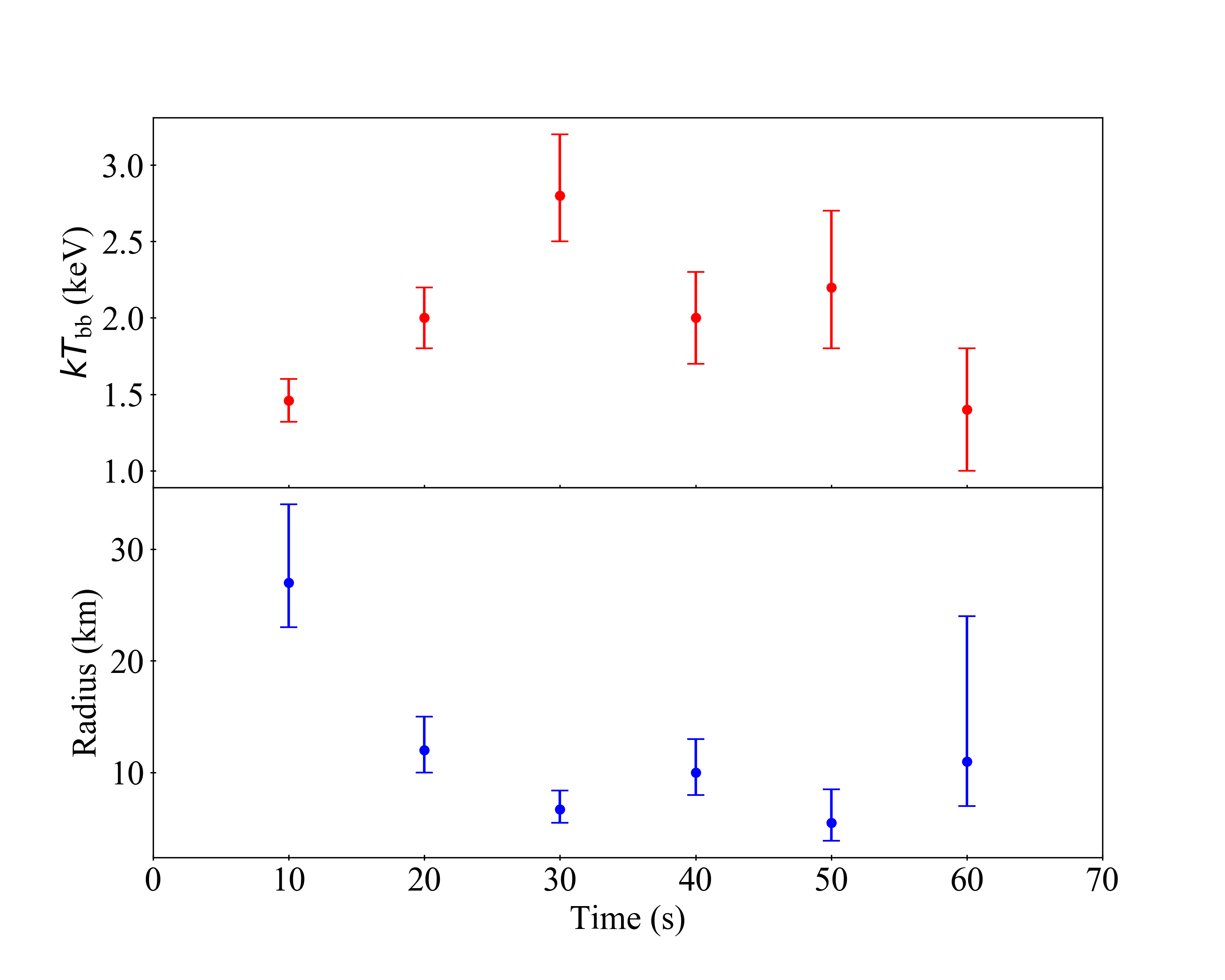}
\includegraphics[height=6 cm, width=8.5 cm]{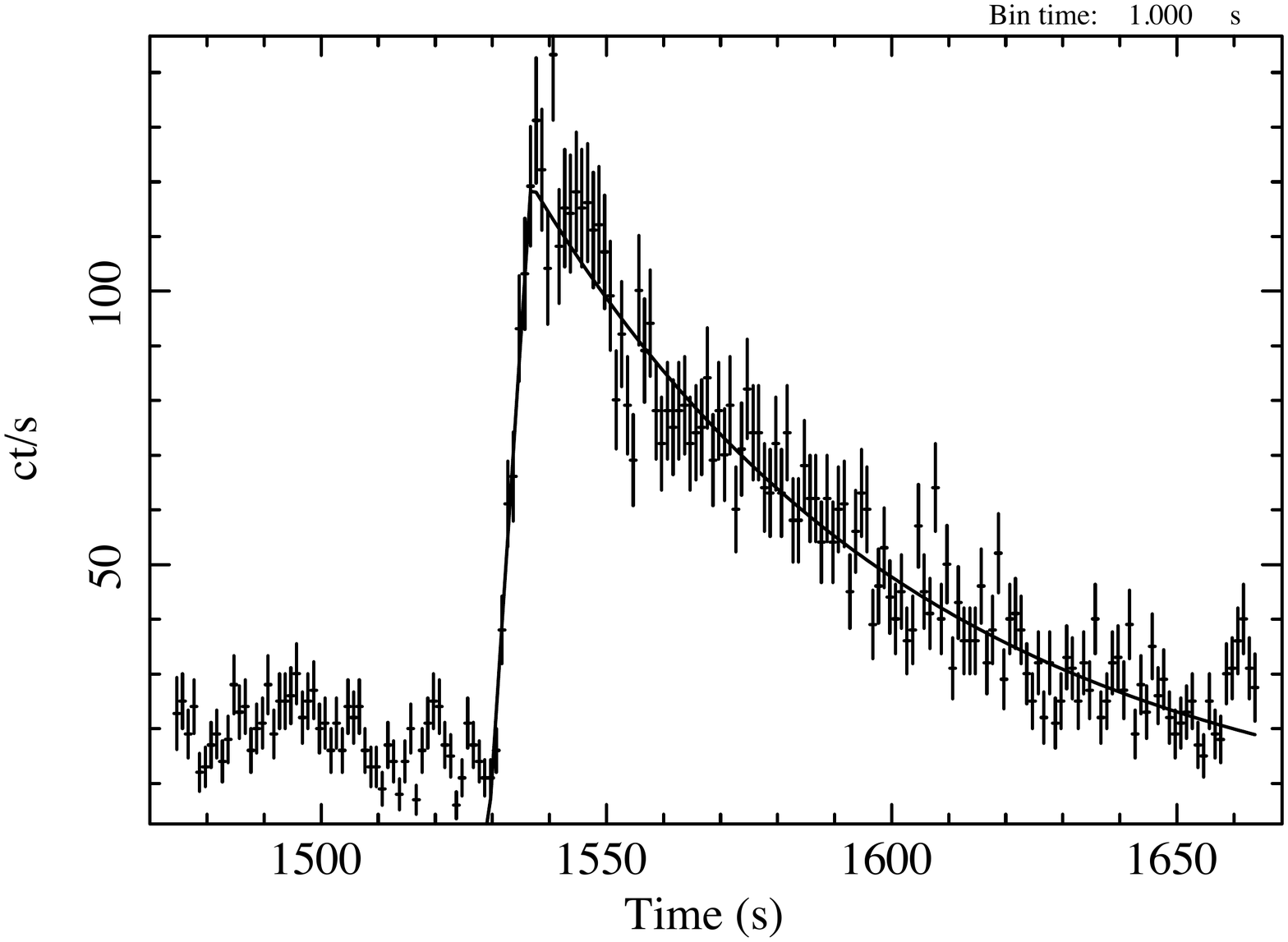}
\includegraphics[height=6.5 cm, width=8 cm]{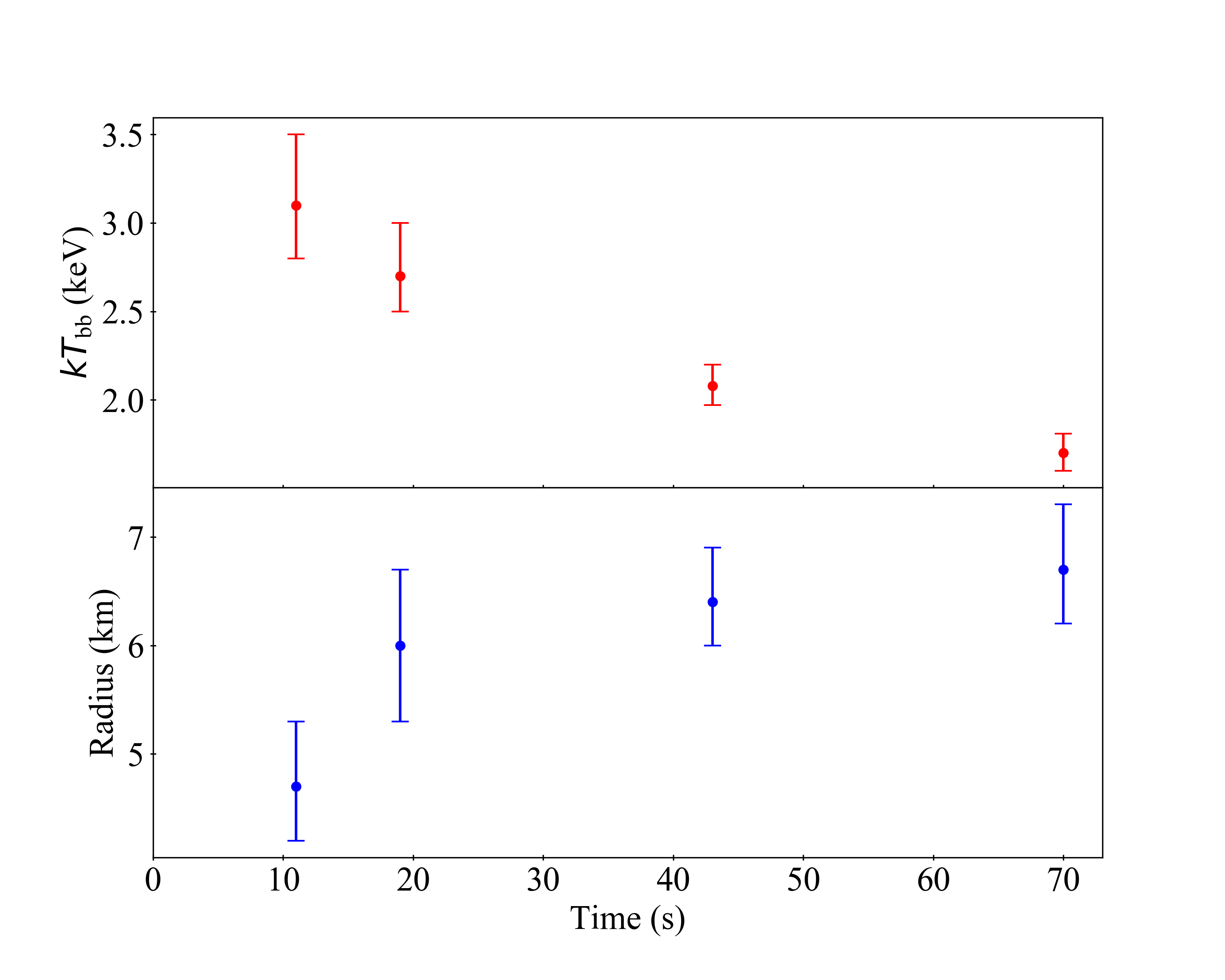}
\includegraphics[height=6 cm, width=8.5 cm]{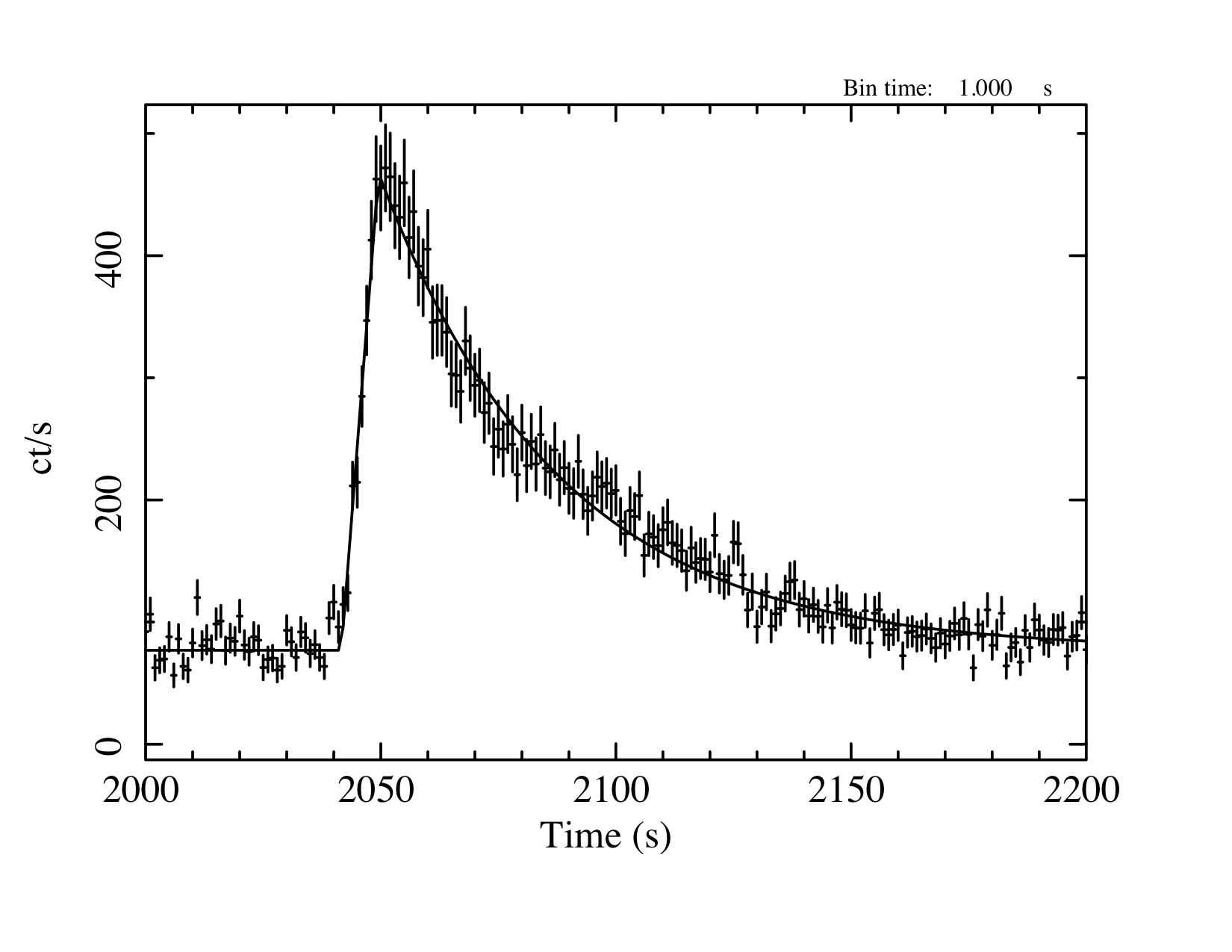}
\includegraphics[height=6.5 cm, width=8 cm]{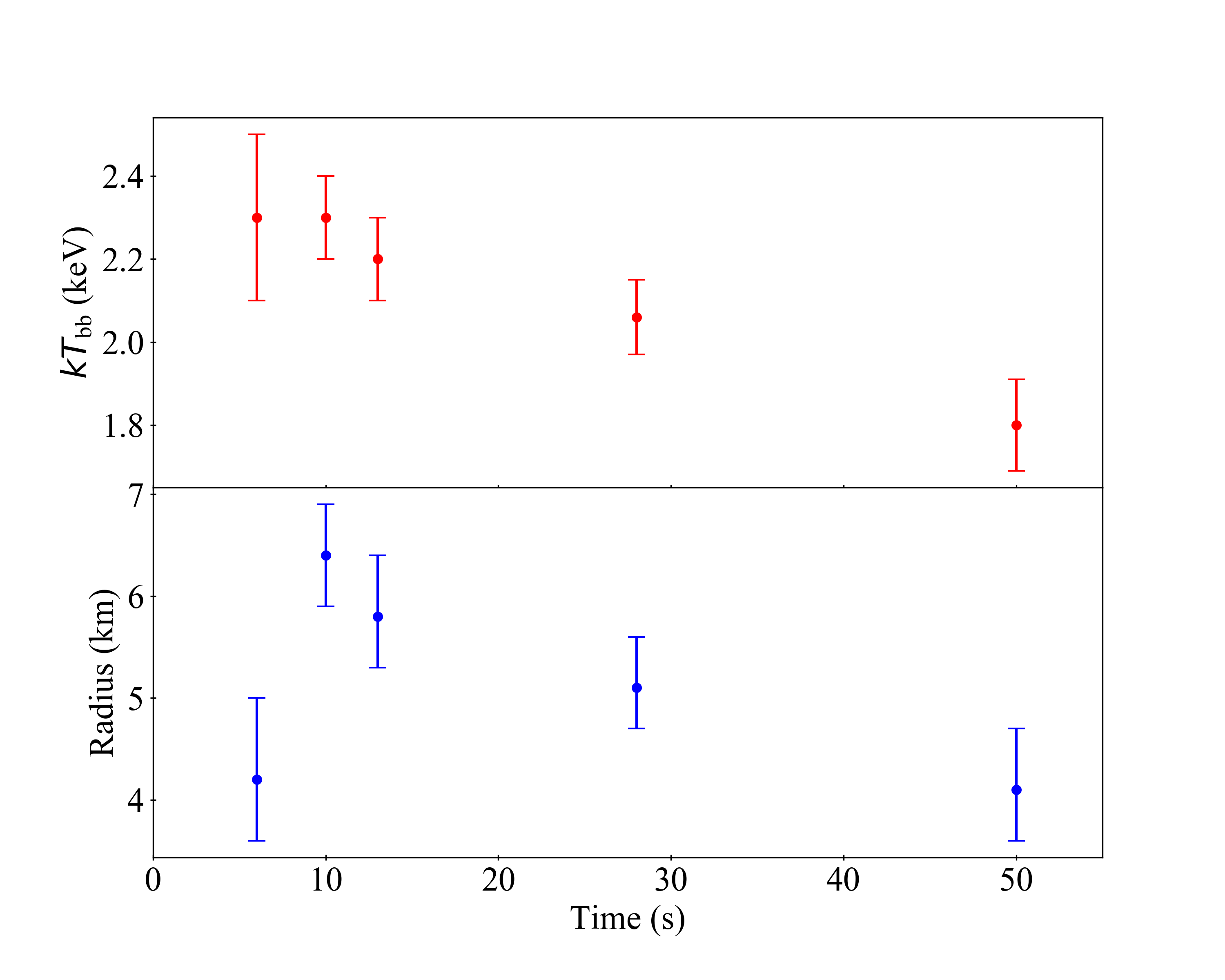}
\caption{Light curve (\emph{left}) and spectral parameters evolution over the whole burst duration (\emph{right}) for the B1 ({\it top}), the B2 ({\it middle}) and the B3 bursts ({\it bottom}). Light curves have been fit by a \textsc{burs} model. B1 is binned at 2 s, while B2 and B3 are binned at 1 s. The zero in the time interval was set to be equal to MJD56033.35963658 for B1, MJD57059.8920679 for B2 and MJD57113.70065027 for B3.} 
\label{fig:spe_evo}
\end{figure*}

We have analysed the 2012 type-I burst observed by JEM-X \citep{Chenevez2012atel} during the very faint emission period of RX1804, five bursts we found during the hard state within the XRT pointings, including the ones reported by \citet{Krimm2015b} and \citet{IntZand2019}, and ten bursts observed by \nustar\ /FPMA during the intermediate spectral state. Each burst profile was fit with the \textsc{burs} model included in the  \textsc{qdp} tool, which well describes their Fast Rise Exponential Decay (FRED) shape. The duration of each burst lasts between 100 and 150 s. We found that the XRT bursts show similar characteristics in terms of peak luminosity, decay time \citep[as in agreement with ][]{IntZand2019} and peak occurrence time, so that in the following we show results only for one of the sample (B2 in Tab. \ref{tab:burst12}). The bursts observed by \nustar\ during the intermediate state are even more peculiar: they occur regularly every $\sim$4 ks with almost the same FRED profile (mean decay time of $\sim$36 s, with a standard deviation of $\sim$2 s). This behaviour is similar to the one displayed by other clocked bursters, the most famous being the "text-book" burster GS 1826--24, which has been regularly exhibiting bursts every $\sim$6 hr for at least 20 years \citep{Ubertini1999,Chenevez2016}. In order to confirm the regularity of the burst recurrence, which can be inferred by eye from Figure \ref{fig:clock} (\emph{Left}), we ran the \textsc{Xronos} tool \textsc{efsearch}, which performs an e-folding of the light curves. We chose a 16-channels binning using 
a trial period of around 4000 s. Then we performed a gaussian fit finding a maximum at 3860 s with a FWHM of $\sim$140 s  (see Figure \ref{fig:clock}, \emph{right}). We conclude then that during the intermediate state the source did display quasi-periodic bursts, a behaviour which is thought to be related to a constant mass-accretion rate \citep{Galloway2004}. Finally, we chose one of the bursts to be representative of the sample of \nustar\ bursts, which in the following will be identified as B3 (see Table \ref{tab:burst12} and Figure \ref{fig:spe_evo}).

\begin{figure}
\centering
\includegraphics[height=5 cm, width=7.5 cm]{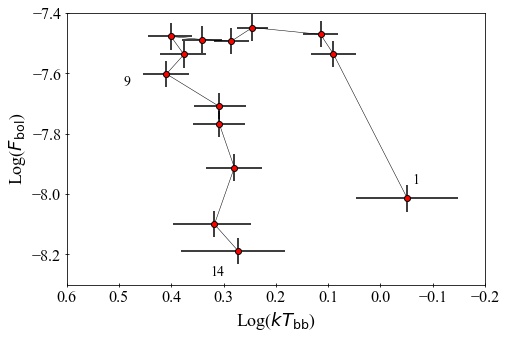}
\caption{Flux-temperature diagram for B1, with a 14-intervals time subdivision. The start time is the same as interval 1 in Table \ref{tab:burst12}. The data points have been connected to give chronological order.}
\label{fig:PRE}
\end{figure}

Despite of the bursts observed by \swift\ and \nustar\ in 2015, the JEM-X burst (B1) shows different characteristics. In particular, in spite of the B1 light curve, the B2 and B3 profiles are well compatible with a standard FRED, without evidence of any  wide structure in the peak, which is evident in B1 (Fig. \ref{fig:spe_evo}, \emph{top}). This could be considered a hint for a double peak, even though the bad statistics does not allow us to carry out a clear profile.\\
Thus, we performed a time-resolved spectral analysis of the three bursts. For B1, we extracted spectra in six time intervals, 10-s long, starting at 8:37:38 UTC, while for B2 and B3 we extracted four and five intervals, respectively of variable duration based on the statistics (see Table \ref{tab:burst12}). For both B2 and B3, in order to neglect the contribution by the persistent emission, we used a spectrum taken $\sim$50 s before the start of each of the two bursts as the background spectrum. We did not apply a similar procedure for B1 since the persistent emission was below the telescope sensitivity. We then fit spectra by using a simple black body model (\textsc{bbodyrad} in \textsc{xspec}), multiplied by the \textsc{tbabs} model. The interstellar equivalent absorption column $N_{\rm H}$ value was fixed at 0.22$\times$10$^{22}$ cm$^{-2}$. In B1, a simultaneous radius increasing and temperature decreasing can be observed around 40 s from the start of the burst, i.e. between the third and the fourth time segment (Tab. \ref{tab:burst12} and Fig. \ref{fig:spe_evo}, \emph{top}). On the contrary, in B2 and B3, the temperature evolution (see Figure \ref{fig:spe_evo}, \emph{middle} and \emph{bottom}) suggests a simple cooling of the NS after the peak of the burst. \\ 
The bolometric unabsorbed flux at the burst peak is 3.2$^{+1.5}_{-1.1}$ $\times$10$^{-8}$ erg cm$^{-2}$ s$^{-1}$ in B1, 2.0$^{+0.5}_{-0.4}$ $\times$10$^{-8}$ erg cm$^{-2}$ s$^{-1}$ in B2 and 1.3$\pm$0.2 $\times$10$^{-8}$ erg cm$^{-2}$ s$^{-1}$ in B3. The flux errors have been derived by fixing the normalization factor of \textsc{bbodyrad} to its lower and upper limits. Then, considering that the normalization $K$ of \textsc{bbodyrad} is connected to the black-body radius $R$ through the relation:  $K = R^2_{\rm km}/D^2_{\rm 10}$, where $D_{\rm 10}$ is the source distance in units of 10 kpc, we computed the apparent radius of the NS for each spectrum (assuming d=10 kpc; see Tab. \ref{tab:burst12}). \\
A further argument in favour of a PRE nature for B1 can be found from the  $F_{\rm bol}$-$kT_{\rm bb}$ diagram throughout the burst, as in \cite{VanParadijs1990} and \cite{Lewin1993}. In such diagrams, the saturation of the flux during the peak is witnessed by the $F_{\rm bol}$ values following an horizontal trend, the so-called "expansion track". In order to have more points to trace the evolution of the burst in the diagram, we repeated the time resolved spectroscopy analysis with a finer time subdivision, i.e. of 14 5-seconds long intervals. In each of these intervals we calculated the 0.1-30 keV unabsorbed flux. As expected in the PRE case, the flux evolution is compatible with a constant up to a certain point (segment 8 in Fig. \ref{fig:PRE}). Thereafter, the cooling phase, as witnessed by the flux decreasing, starts (9 in the same figure).\\
Finally, PRE bursts are more typically observed at low accretion regimes with He-rich material, when the fuel is more efficiently stored onto the NS surface, e.g. as in 4U 1812-12 \citep{Cocchi2000}. The 2015 bursts (including B2 and B3), on the contrary, are observed with likely short recurrence times in outburst, with duration and profiles fully compatible with the ignition of H-rich material, a trend which reminds of the "clocked burster" GS 1826-238 \citep[see, e.g. ][]{Zamfir2012, Galloway2004}, where PRE bursts are never observed. The 2012 burst (B1) occurred on the contrary during a low-level accretion phase and it might be more suggestive of a PRE episode.\\

\begin{table*}
\centering
\begin{tabular}{l l l l l l l l l l l l l l}
\hline 
\hline
\multicolumn{13}{c}{{\bf General properties of the bursts}}\\
\hline
& & \multicolumn{3}{c}{B1} & & \multicolumn{3}{c}{B2} & & \multicolumn{3}{c}{B3} \\
\cmidrule(lr){3-5} \cmidrule(lr){7-9} 
\cmidrule(lr){11-13}
Instrument & & \multicolumn{3}{l}{JEM-X (3-20 keV)} & & \multicolumn{3}{l}{XRT (0.5-10 keV)} & &  \multicolumn{3}{l}{\nustar\ (3-50 keV)} \\
Occurrence Time (MJD) & & \multicolumn{3}{l}{56033.35986} & & \multicolumn{3}{l}{57059.89270} & & \multicolumn{3}{l}{57113.70346} \\
Peak Time (s) & & \multicolumn{3}{l}{17.0$^{+3.0}_{-2.0}$} & & \multicolumn{3}{l}{7.7$\pm$0.5} & & \multicolumn{3}{l}{6.9$\pm$0.5}\\
Decay Time (s) & & \multicolumn{3}{l}{24.0$^{+5.0}_{-4.0}$} & & \multicolumn{3}{l}{67.0$\pm$3.0} & & \multicolumn{3}{l}{38.0$\pm$5.0} \\
Peak count rate (cts/s) & & \multicolumn{3}{l}{ 270$\pm$40} & & \multicolumn{3}{l}{121$^{+3}_{-4}$} & & \multicolumn{3}{l}{391$^{+5}_{-4}$}\B\\
\hline  
\multicolumn{13}{c}{{\bf Time Resolved Spectroscopy}}\\
\hline
{Interval} & & Duration & {$kT_{\rm bb}$} & {Radius} & & {Duration} & {$kT_{\rm bb}$} & {Radius} & & Duration & {$kT_{\rm bb}$} & {Radius}  \\
 & & s & keV & km  & & s & keV & km  \\
\hline
{1} & & 10 & {1.46$\pm$0.14} & {27.0$^{+7.0}_{-4.0}$} & & 11.00 & {3.1$^{+0.4}_{-0.3}$} & {$4.7^{+0.6}_{-0.5}$} & & 6.00 & 2.3$\pm$0.2 & 4.2$^{+0.8}_{-0.6}$\\
{2} & & 10 & {2.00$^\pm$0.20}  & {12.0$^{+3.0}_{-2.0}$} & &  8.10 & {2.7$^{+0.3}_{-0.2}$}  & {6.0$\pm$0.7} & & 10.00 & 2.3$\pm$0.1 & 6.4$\pm$0.5 \\
{3} & & 10 & {2.80$^{+0.40}_{-0.30}$} & {6.7$^{+1.7}_{-1.2}$} & &  24.18 & {2.08$^{+0.12}_{-0.11}$} & {6.4$^{+0.5}_{-0.4}$} & & 13.00 & 2.2$\pm$0.1 & 5.8$^{+0.6}_{-0.5}$\\
{4} & & 10 & {2.00$\pm$0.30} & {10.0$^{+3.0}_{-2.0}$} & & 27.0 & {1.70$^{+0.11}_{-0.10}$} & {6.7$^{+0.6}_{-0.5}$} & & 28.00 & 2.06$\pm$0.09 & 5.1$^{+0.5}_{-0.4}$ \\
\cmidrule(lr){7-9}
{5} & & 10 & {2.20$^{+0.5}_{-0.4}$} &{5.5$^{+3.0}_{-1.6}$}  & & & & & & 50.00 & 1.80$\pm$0.11 & 4.1$^{+0.6}_{-0.5}$ \\
\cmidrule(lr){11-13}
{6} & & 10 & {1.40$\pm$0.40} & {11.0$^{+13.0}_{-4.0}$}\\
\hline
\hline
\end{tabular}
\caption{Summary of the results obtained from the three selected yype-I X-ray bursts, through light curves analysis and time resolved spectroscopy. The decay time is the characteristic time interval for the count-rate to decay by a factor 1/e. In the spectral results for B1 we used JEM-X1 and JEM-X2 combined data, while XRT data were used for B2 and \nustar\ data for B3. Spectra are fitted with a \textsc{tbabs}$\times$\textsc{bbodyrad} model. Errors at 68\% confidence level are reported.}
\label{tab:burst12}
\end{table*}

\section{Discussion}
We have studied the NS LMXB RX1804 in two different accretion-rate phases: in 2015, when the source performed a 4-month long outburst, with a peak luminosity of $\sim$2$\times$ 10$^{38}$ \ergs\ , i.e. $\sim$0.5 $L_{\rm Edd}$ and in 2012, when the persistent luminosity of the outburst was a few 10$^{35}$ \ergs\ , implying a VFXT phase. \\ NS-LMXBs displaying both very faint and bright outbursts have been classified as "hybrid" sources \citep{Degenaar2009_vfxt}. The physical mechanisms allowing these systems to display such different outbursts luminosities are still unknown. However, it is interesting to notice that this behaviour is somewhat similar to the BH transients showing outbursts at peak luminosities spanning a range of two-three orders of magnitude, but almost always above the detection threshold of the large field of view X-rays telescopes ($\sim$10$^{36}$ erg s$^{-1}$). Since NS-LMXBs are intrinsically fainter than BH transients, an analogous peak luminosity range for this class would occur at the turn of this detection threshold. Thus, we suggest that the unknown mechanism for hybrid outbursts in NS-LMXBs could be the same mechanism leading BH transients to show different peak luminosities, which is similarly unclear. \\
In the following we discuss the main results reported in this paper.

\subsection{The spectral evolution of the source in the 2015 outburst: evidence of an intermediate state}
\begin{figure}
\centering
\includegraphics[height=6 cm, width=9 cm]{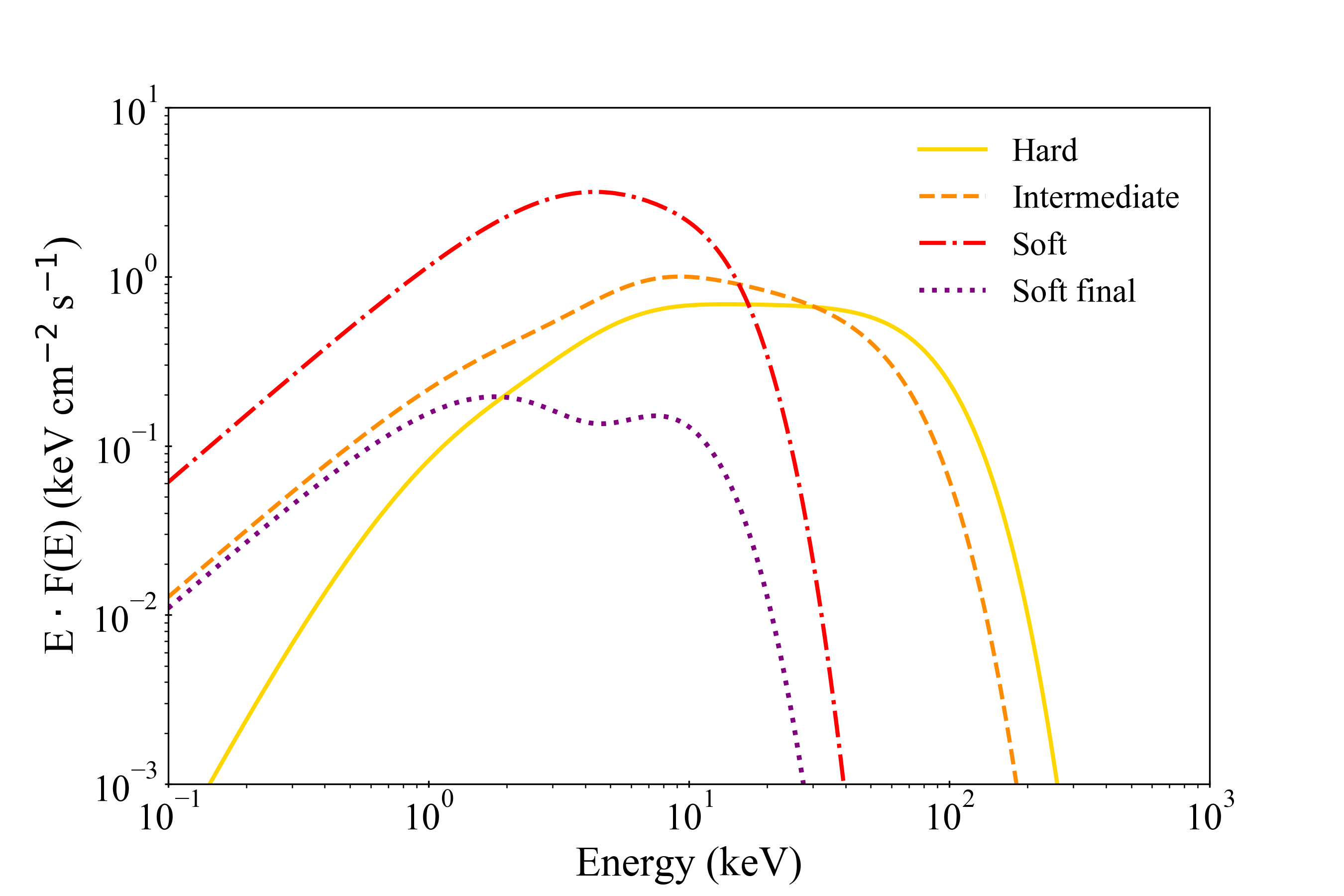}
\caption{Best-fit models of observations H1 (yellow continuous line), I1 (orange dashed line), S4 (red dot-dashed line) and S5 (violet dotted line).} 
\label{fig:models}
\end{figure}

The \swift\ spectral analysis highlights how the RX1804 spectrum evolves during the outburst. In Fig. \ref{fig:models} we show the best-fit models of the system in four phases of its evolution throughout the outburst, which displays four different spectral states. The hard state spectra H1 and H2, collected during the "plateau" in the light curve, are well described by a double-seeds Comptonization spectrum, arising from the interaction of the spectra by two different photon sources with the same hot electron plasma. Indeed, as suggested by \cite{Cocchi2011}, NS-LMXBs in the hard state might be divided in two sub-classes: the one-photon population (1P) and the two-photon population (2P). In the former we can find only one source of photons,  which can be the NS or the boundary layer, Comptonized by a hot corona. In the 2P systems also the photons emitted by the disk are Comptonized by the same medium, leading to a spectrum consisting of two different thermal Comptonization spectra. This is most likely related to the accretion luminosity observed in the hard state of 2P population, which is consistently higher than the hard state luminosity of the 1P sources \citep{Cocchi2011}. Indeed, RX1804 shows a luminosity in hard state of $\sim$4$\times 10^{37}$ \ergs\ . \\ We suggest here that the so-called "very hard" spectral state observed in this source by \cite{Parikh2017_vhard} might be explained, at least in the case of RX1804, in the framework of a thermal Comptonization spectrum with two seed photons populations. Indeed, for a corona temperature typical of NS-LMXBs in hard state, the presence of two Comptonization spectra with different seed photons makes the soft X-rays spectrum particularly flat, in particular with respect to a single Comptonization spectrum. Broad band X-rays spectral studies might point out whether the same explanation works for any of the other claimed "very hard" sources \citep{Wijnands2015,Parikh2017_vhard,Wijnands2017_vhard}, with the exception of IGR J17361-4441, which was well-modeled by a single Comptonization spectrum \citep{DelSanto2014}. However, it is worth noting that the X-ray binary nature for IGR J17361-4441 is far from certain, as suggested by \cite{DelSanto2014} and \cite{Bozzo2014}. \\
The corona electron temperature does not change between H1 and H2 (leveling off at a value around 18 keV), while a significant corona cooling is observed in I1, $\sim$12 keV. This phenomenon, combined with the arising of the direct disk emission in the data, points out that the source was in an intermediate state, which is quite common in black hole X-ray binaries, but rarely observed in NS-LMXBs. It is remarkable that the change in the spectral shape between the hard state and the intermediate happens right after the "peak" in the BAT light curve, only up to 40 keV (see Figure \ref{fig:swift}). This is consistent with the pivot at 40 keV present between the hard and the intermediate state (see Fig. \ref{fig:models}) due to the receding of the spectral cut-off, corresponding to the corona electron temperature decrease.
In the soft state spectra from S1 to S4, we observe steady values of the temperatures of both disk and NS, except for S5, where the disk is remarkably colder. This is coherent with the outburst fading and the disk depletion, witnessed by the increase of the inner disk radius from S4 to S5. Including \nustar\ data in S4, we are able to fit the spectrum with a more physically motivated model which replaces the \textsc{bbody} model with a \textsc{comptb} with $\tau \approx$ 11 (see Table \ref{tab:refl}). As expected, the black body model used in the soft state for the \swift\ spectra is only a crude approximation of a saturated Comptonization spectrum.\\
The discussed evolution of the spectral parameters of RX1804 throughout the outburst, in particular the increasing of the disk temperature, combined with the corona cooling, can be explained in the framework of a disk truncated at large gravitational radii in hard state and then approaching towards the compact object over the transition to the intermediate and soft states \citep[e.g., ][]{Barret2002,Dai2010,Gambino2019}. The alternative plane-parallel corona scenario, which assumes an extended disk close to the NS surrounded by an optically thick corona in hard state and decreasing progressively in size going through the intermediate and soft states ("compactification"), seems unlikely since the relatively low $\tau$ values we have found.  

\subsection{The reflection spectrum and the hard tail with \nustar}
The analysis carried out with \nustar\ provides the evidence of a low reflection fraction (<20\%) and (as  in agreement with \cite{Ludlam2016,Degenaar2016}) a relatively high ionization parameter, in all the three spectral states of RX1804. On the contrary, we were not able to give constraints on the inclination of the system, which was estimated with the same \nustar\ spectra as $\sim$18-29$^\circ$ in the hard state \citep{Ludlam2016} and as $\sim$27-35$^\circ$ in the soft state \citep{Degenaar2016}, thus we estimated the spectral parameters assuming two different cases, i.e. for a low (20$^\circ$) and high (60$^\circ$) inclination. We suggest that this inconsistency with the previous results is possibly due to the different spectral binning we chose (see Subsection \ref{ss:nustar}). \\ The fits result in wide inner disk radius ranges, which include the estimates found with the \textsc{diskbb} model (this work) and the results by \cite{Ludlam2016} and \cite{Degenaar2016}. \\ It is worth noticing that our bolometric flux estimations for H1 and S4 are at least twice the values reported by \cite{Ludlam2016} and \cite{Degenaar2016}. We suggest that this discrepancy is likely due to the procedure of co-adding the spectra, the background files and the ancillary response files using \textsc{addascaspec}. \\
Indeed we tried to sum FPMA and FPMB spectra in the hard state with \textsc{addascaspec} obtaining a 1.9$\times$10$^{-9}$ \ergcm flux in the 0.45-50 keV energy range, which is compatible within 10\% with the estimate found by these authors. We verified that in our case the correct flux value is obtained by using \textsc{addascaspec} with the keyword \texttt{errmeth=GAUSS}, or using the \textsc{specadd} tool.\\ 
Unlike the \nustar\ study in the hard and soft states reported in the previously cited papers, we report here on the first broad band analysis of RX1804 including \nustar\ data in intermediate state. The physical parameters of the accretion found in the \nustar-including spectrum are compatible with the results by the \swift-only, confirming the evidence of the system being in a (rarely observed) intermediate state. The difficulty of catching intermediate spectral states in NS-LMXBs, witnessed by the very small number of reported intermediate states in literature, might be due to very rapid hard-to-soft transitions in this class of systems, significantly faster than the analogous transitions in BH X-ray binaries \citep[as highlighted by ][]{MunozDarias2014}, since the presence of the NS surface makes cooling more efficient. \\
The inclusion of \nustar\ analysis led us to individuate an additional power-law contribution in all the three states displayed by RX1804. In all three spectra, indeed, including a power-law results in a statistically significant improvement of the fit. Unfortunately, only in one case, i.e. I1, the introduction of the additional component resulted in a stable fit with a flux contribution to the total of $\sim$50\% in the 30-100 keV energy band. It is noteworthy that the fit in the H1 spectrum gives odd results (i.e. a $kT_{\rm s,1}$ of 3-4 keV) and rather unstable fits even including a power-law component. We suggest that the simple prescription of a double-seeds Comptonization scenario with two seed photons populations scenario with a single $kT_{\rm e}$ and $\tau$ for the hot corona might be either an oversimplification in this case, or that a spectral model capable to take into account properly both a thermal and non-thermal populations of photons is needed. \\
With the advent of {\it{RXTE}}, {\it{BeppoSAX}}, and later \integral, hard tails have been observed in the brightest NS-LMXBs, the so-called Z-sources \citep[][ and reference therein]{DiSalvo2006}, and in the bright Atoll-sources \citep{Paizis2006}. The formers show very soft spectra modelled with a black-body plus a thermal Comptonization at few keV and a variable power-law component above 30 keV which contributed from 1\% to 10\% to the source luminosity \citep{DiSalvo2000,Iaria2001,DiSalvo2001}. On the contrary bright-Atoll sources, show dramatic changes in the spectral state, and in a few cases hard tails have been detected during the hard state \citep{Tarana2011}, intermediate states
\citep{Fiocchi2006,Paizis2006, Tarana2007} and soft state \citep{Piraino2007,Pintore2015}. These results might include RX1804 in the sample of the bright-Atolls showing hard tails in addition to the thermal Comptonization spectrum.\\
The nature of this component is still debated as that observed in BH binaries, both in soft and intermediate state \citep{, Malzac2006,DelSanto2008}, and more recently also in hard state \citep{Bouchet2009,DelSanto2013}. The most accepted explanation is that a hybrid thermal/non-thermal population of electrons co-exist in the Comptonizing corona, just causing the non-thermal power-law component observed at higher energy \citep{Poutanen1998, Coppi1999}. 

\subsection{Distance of RX1804 and its burst phenomenology}
According to the time resolved spectral analysis (and the other clues presented in Section \ref{sec:burst}) of the 2012 type-I burst, it results that it underwent most likely a photospheric radius expansion, as assumed by \cite{Chenevez2012atel} and \cite{Chelovekov2017}. Thus, the distance of RX1804 can be easily derived by using the critical Eddington luminosity reported in \cite{Kuulkers2003}, i.e. $\simeq$ 3.79$\times$10$^{38}$ erg s$^{-1}$ with a 4\% uncertainty. With a peak flux of 3.2$^{+1.5}_{-1.1}\times$10$^{-8}$ erg cm$^{-2}$ s$^{-1}$, we obtain a distance of 10.0$^{+4.7}_{-3.5}$ kpc. The wide error range for the distance is a consequence of our conservative choice to estimate the errors of the peak flux (see subsection \ref{sec:burst}). If, alternatively, we consider a standard 10\% uncertainty, we find a 10.0$\pm$1.4 kpc distance, which is compatible with the upper limit value reported by \cite{Chelovekov2017}. This is significantly discrepant with the value of 5.8 kpc estimated by \cite{Chenevez2012atel}. The reason is likely to rely on the method these authors used to estimate the burst peak flux, i.e. they compared the peak count-rate with the count-rate of the Crab, in order to express the flux in crab units, and then converting this flux in c.g.s. This method can be reliable only when sources show Crab-like spectra or when fluxes are calculated within small energy bands. However, this is not the case for a type-I X-ray burst, whose spectrum has almost a black-body spectral shape, while the Crab spectrum is described by a power-law with $\Gamma$ index of 2.1. \\
The relatively high distance found for RX1804 would put the system likely beyond the Galactic Center. This is not troublesome, since the system has a $\sim$6.6$^\circ$ angular separation from Sgr A* and it is probably located in or right beyond the Galactic Bulge. This region seems to be inhabited by most of the VFXTs discovered so far \citep{Intzand2001,Muno2005,Wijnands2006_GC,Degenaar2012_GC}, therefore this new estimate of the distance seems reasonable. However, although using PRE bursts to infer distances is an extremely useful tool and it has been vastly employed to NS-LMXBs in the past, the method does suffer from systematic uncertainties. Indeed, not all the bursts showing PRE from one source reach the same peak flux, but they rather scatter around a mean value, with variations usually within 15\% \citep{Kuulkers2003,Galloway2003,Galloway2008}. Furthermore, without precise information on the mass and the radius of the NS or on its atmospheric composition, the value used for the Eddington luminosity may not be accurate for the source. It is therefore necessary to consider the distance estimates given in this paper with a bit of caution. \\ The average duration of all the bursts reported in this work, i.e. around 100 s, seems compatible with low mass-accretion rates \citep{Fujimoto1987} of a H/He mixed material, which usually gives bursts lasting less than 100 s \citep{Intzand2005}. Furthermore, the "clocked burster" behaviour requires also accreted material with low metallicity \citep{Galloway2004, Lampe2016}. Although its burst phenomenology likely points out the presence of H and He in the accreted material, it is noteworthy that its optical spectrum seems somewhat lacking of hydrogen and He I lines, as reported by \cite{Baglio2016}. This evidence, combined with the detection of a He II line, was used by the authors to propose RX1804 being an UXCB harbouring a helium white dwarf.However, a helium dwarf companion scenario seems hard to reconcile with the hydrogen/helium bursts exhibited by RX1804, therefore the UCXB nature of the system needs additional evidence to be confirmed. \\ 
We propose, as alternative explanation, that RX1804 harbours a brown dwarf (BD) companion, as in the case of SAX J1808.4--3658 \citep{Bildsten2001}, which also shows bursts from mixed H/He material \citep{Galloway2006_1808}. In addition, BDs have been proposed as donor stars in VFXT systems by \cite{King2006}. However, it is worth noticing that hydrogen lines are present in the optical spectrum of SAX J1808.4--3658 \citep{Elebert2009}. This difference between the two sources might be due to either the sensitivity of the optical telescope used, the lower distance of SAX J1808.4--3658 \citep{Galloway2006_1808}, or both.
\section{Summary \& Conclusions}
In this work we attempted to follow a comprehensive study of RX1804, an ex-"burst only" source, in two aspects of its X-ray activity: the long outburst occurred in 2015 and type-I X-ray burst. We tried first to reconstruct the spectral behavior of the source during the whole 4-months outburst taking advantage of the broad band coverage by XRT and BAT together. We then paired three \nustar\ observations with three, quasi-simultaneous, XRT and BAT spectra to analyze the reflection continuum. On the other hand, we performed time resolved spectroscopy of three bursts, one observed by \integral\, one by XRT during the hard state and one by \nustar\ when the source was in intermediate state. Our results can be summarized as follows:
\begin{itemize}
    \item we were able to follow the spectral evolution of the source all along the outburst, catching the source in an intermediate spectral state, rarely observed in NS-LMXBs.
    \item the \nustar\ and BAT data reveal the arising of a hard tail in the intermediate state, modelled as a power-law, pointing out that non-thermal processes are likely at play during the state transition.
    \item in the hard state, a model composed of two Comptonization spectra interacting with the same hot corona well describes the data, indicating that RX1804 belongs to the "two-photon" population, as e.g. the other bright burster GS 1826-238 \citep{Cocchi2011}. We suggest here that the presence of two Comptonization spectra does explain the so-called "very hard" spectral state claimed for the source \citep{Parikh2017_vhard}.
    \item the \nustar\ analysis confirms the accretion physics scenarios resulted by the \swift\ spectra. In addition, it reveals a reflection contribution below 20\% in all the states. However, with our Compton reflection modeling, we were able to provide only lower limits on the inner disk radius and no constraints on the inclination of the system. 
    \item we found different hints suggesting that a Photospheric Radius Expansion burst occurred in 2012. This allowed us to  confirm the distance of $\sim$10 kpc  estimated by \cite{Chelovekov2017}. 
    \item interestingly, during the intermediate, we unveiled a "clocked burster" behaviour, with recurrence time of roughly 4000 s.
\end{itemize}
It is still unclear how systems like RX1804 exhibit different levels of X-ray activity in time-scales of years, transitioning from very faint outbursts to bright outbursts, as also observed in BH transients. This peculiar (and quite common) behaviour remains an open question in the framework of the Disk Instability Model, shedding light on how important and necessary are further investigations on the topic. 

\section*{Acknowledgements}
Authors acknowledge financial contribution from the agreement ASI-INAF n.2017-14-H.0, from INAF mainstream (PI: T. Belloni) and from the HERMES project financed by the Italian Space Agency (ASI) Agreement n. 2016/13 U.O. The research leading to these results has received funding from the European Union’s Horizon 2020 Programme under the AHEAD project (grant agreement n. 654215). AM acknowledges funding from FSE (Fondo Sociale Europeo) Sicilia 2020. JM acknowleges financial support from PNHE in France and  from the OCEVU Labex (ANR-11-LABX-0060) and the A*MIDEX project (ANR-11-IDEX-0001-02) funded by the "Investissements d’Avenir" French government program managed by the ANR. RI and TDS acknowledge the research grant “iPeska” (PI: Andrea Possenti) funded under the INAF national call Prin-SKA/CTA approved with the Presidential Decree 70/2016. AM thanks Teresa Mineo for useful discussion on the clocked bursts analysis.

\bibliography{biblio}
\end{document}